\documentclass[12pt]{article}

\usepackage{rotating}
\usepackage{sgame}
\usepackage{color}

\def\smallromani{\renewcommand{\theenumi}{\roman{enumi}}
\renewcommand{\labelenumi}{(\theenumi)}}

\newcommand{\Proof}{\NI
                    {\bf Proof.}\ }
\newtheorem{theorem}{Theorem}[section]
\newtheorem{defined}[theorem]{Definition}

\newtheorem{exa}[theorem]{Example}

\newtheorem{lemma}[theorem]{Lemma}
\newtheorem{corollary}[theorem]{Corollary}

\newtheorem{note}[theorem]{Note}

\newtheorem{exe}{Exercise}

\newtheorem{pro}{Problem}

\newcounter{symbol}
\newcommand{\indexsyma}[1]%
{\stepcounter{symbol}\index{zzz1 \thesymbol @\protect#1}}
\newcommand{\indexsymb}[1]%
{\stepcounter{symbol}\index{zzz2 \thesymbol @\protect#1}}
\newcommand{\indexsymc}[1]%
{\stepcounter{symbol}\index{zzz3 \thesymbol @\protect#1}}
\newcommand{\indexsymd}[1]%
{\stepcounter{symbol}\index{zzz4 \thesymbol @\protect#1}}
\newcommand{\indexsyme}[1]%
{\stepcounter{symbol}\index{zzz5 \thesymbol @\protect#1}}

\newcommand{\Searrow}{\begin{turn}{45}$\Downarrow$    \end{turn}}
\newcommand{\Swarrow}{\begin{turn}{135}$\Uparrow$   \end{turn}}

\newcommand{\oldbfe}[1]{\begin{bfseries}\emph{#1}\end{bfseries}}

\newcommand{\ES}{\mbox{$\emptyset$}}

\newcommand{\myra}{\mbox{$\:\rightarrow\:$}}

\newcommand{\Ra}{\mbox{$\:\Rightarrow\:$}}
\newcommand{\tra}{\mbox{$\:\rightarrow^*\:$}}

\newcommand{\sse}{\mbox{$\:\subseteq\:$}}

\newcommand{\LL}{\mbox{$\ldots$}}

\newcommand{\C}[1]{\mbox{$\{{#1}\}$}}           

\newcommand{\NI}{\noindent}
\newcommand{\HB}{\hfill{$\Box$}}
\newcommand{\VV}{\vspace{5 mm}}
\newcommand{\III}{\vspace{3 mm}}
\newcommand{\II}{\vspace{2 mm}}

\newcommand{\szkew}[1]{\relax \setbox0=\hbox{\kern -24pt $\displaystyle#1$\kern 0pt }%
\box0}
{\catcode`\@=11 \global\let\ifjusthvtest@=\iffalse}

\newcounter{oldmycaption}

\usepackage{handbookbib}
\usepackage{amsfonts}
\usepackage{latexsym}
\usepackage{alltt}

\title{Uniform Proofs of Order Independence for Various Strategy Elimination Procedures}

\author{Krzysztof R. Apt \\
\emph{School of Computing, National University of Singapore} \\
\emph{3 Science Drive 2, Republic of Singapore 117543}
\footnote{On leave from CWI, Amsterdam, the Netherlands and
University of Amsterdam.}
}
\begin{document}
\maketitle

\date{}

\begin{abstract}

  We provide elementary and uniform proofs of order independence for various
  strategy elimination procedures for finite strategic games, both for
  dominance by pure and by mixed strategies.  The proofs follow the
  same pattern and focus on the structural properties of the dominance
  relations.  They rely on Newman's Lemma (see \cite{New42}) and
  related results on the abstract reduction systems.
\end{abstract}

\newpage

\contentsline {section}{\numberline {1}Introduction}{3}
\contentsline {subsection}{\numberline {1.1}Preliminaries}{3}
\contentsline {subsection}{\numberline {1.2}Background}{4}
\contentsline {subsection}{\numberline {1.3}Motivation}{6}
\contentsline {subsection}{\numberline {1.4}Organization of the paper}{8}
\contentsline {section}{\numberline {2}Abstract Reduction Systems}{9}
\contentsline {section}{\numberline {3}Dominance Relations}{11}
\contentsline {section}{\numberline {4}Pure Strategies: Inherent Dominance}{16}
\contentsline {section}{\numberline {5}Mixed Dominance Relations}{20}
\contentsline {section}{\numberline {6}Mixed Strategies: Inherent Dominance}{24}
\contentsline {section}{\numberline {7}More on Abstract Reduction Systems}{26}
\contentsline {section}{\numberline {8}Pure Strategies: Payoff Equivalence}{29}
\contentsline {section}{\numberline {9}Mixed Strategies: Randomized Redundance}{30}
\contentsline {section}{\numberline {10}Combining Two Dominance Relations}{32}
\contentsline {section}{\numberline {11}Combining Nice Weak Dominance with Payoff Equivalence}{34}
\contentsline {section}{\numberline {12}Combining Two Mixed Dominance Relations}{39}
\contentsline {section}{\numberline {13}Combining Nice Weak Dominance with Randomized Redundance}{41}
\contentsline {section}{\numberline {14}Conclusions}{45}
\contentsline {section}{References}{47}

\newpage

\section{Introduction}
\label{sec:introduction}

\subsection{Preliminaries}
\label{subsec:prelim}

To properly discuss the background for this research we need to recall
a number of concepts commonly used in the study of strategic games.
We follow here a standard terminology of the game theory, see, e.g.,
\cite{Mye91} or \cite{OR94}.  We stress the fact that we deal here
only with \emph{finite} games.  Given $n$ players we represent a
strategic game by a sequence
\[
(S_1, \LL, S_n, p_1, \LL, p_n),
\] 
where for each $i \in [1..n]$

\begin{itemize}
\item $S_i$ is the \emph{finite}, non-empty, set of \oldbfe{strategies} 
(sometimes called \oldbfe{pure strategies})
available to player $i$,

\item $p_i$ is the payoff function for the  player $i$, so

\[
p_i : S_1 \times \LL \times S_n \myra \cal{R},
\]
where $\cal{R}$ is the set of real numbers.
\end{itemize}
We assume that $S_i \cap S_j= \ES$ for $i \neq j$.

Given a sequence of non-empty sets of strategies $S_1, \LL, S_n$ and
$s \in S_1 \times \LL \times S_n$ we denote the $i$th element of $s$ by $s_i$ and
use the following standard notation:

\begin{itemize}
\item $s_{-i} := (s_1, \LL, s_{i-1}, s_{i+1}, \LL, s_n)$,

\item $(s'_i, s_{-i}) := (s_1, \LL, s_{i-1}, s'_i, s_{i+1}, \LL, s_n)$, where
we assume that $s'_i \in S_i$.
In particular $(s_i, s_{-i}) = s$,

\item $S_{-i} := S_1 \times \LL \times S_{i-1} \times S_{i+1} \times \LL \times S_n$,

\item $(S'_i, S_{-i}) := S_1 \times \LL \times S_{i-1} \times S'_i \times S_{i+1} \times \LL \times S_n$.

\end{itemize}
We denote the strategies of player $i$ by $s_i$, possibly with some superscripts.

Next, given a game $G := (S_1, \LL, S_n,$ $p_1, \LL, p_n)$ and non-empty sets of
strategies $S'_1, S''_1, \LL, S'_n, S''_n$ such that $S'_i \sse S_i$
and $S''_i \sse S_i$ for $i \in [1..n]$ we say that $G' := (S'_1, \LL,
S'_n, p_1, \LL, p_n)$ and $G'' := (S''_1, \LL, S''_n, p_1, \LL, p_n)$
are \oldbfe{restrictions}\footnote{Sometimes the name \oldbfe{reduction} is used. In \cite{GKZ90}
a restriction is called a \oldbfe{subgame}.}
of $G$ and denote by $G' \cap G''$ the restriction
$(S'_1 \cap S''_1, \LL, S'_n \cap S''_n, p_1, \LL, p_n)$.  In each
case we identify each payoff function $p_i$ with its restriction to
the Cartesian product of the new strategy sets.

Fix a game $(S_1, \LL, S_n, p_1, \LL, p_n)$.  
We now introduce a number of well-known binary dominance
relations on strategies.
We say that
a strategy $s_i$ is \oldbfe{weakly} (\oldbfe{strictly}) \oldbfe{dominated}
by a strategy $s'_i$, or equivalently, 
a strategy $s'_i$ \oldbfe{weakly} (\oldbfe{strictly}) \oldbfe{dominates} a strategy $s_i$, if

\[
p_i(s_i, s_{-i}) \leq p_i(s'_i, s_{-i})
\]
for all $s_{-i} \in S_{-i}$, with some inequality (all inequalities)
being strict.
We denote the weak dominance relation by $W$ and the strict dominance relation by $S$.

Further, we say that the strategies $s_i$ and $s'_i$ of player $i$
are \oldbfe{compatible} if for all $j \in [1..n]$ and 
$s_{-i} \in S_{-i}$
\[
\mbox{$p_{i}(s_i, s_{-i}) = p_{i}(s'_i, s_{-i})$ implies $p_{j}(s_i, s_{-i}) = p_{j}(s'_i, s_{-i})$.}
\]
We then say that $s_i$ is \oldbfe{nicely weakly dominated by}
$s'_i$ if $s_i$ is weakly
dominated by $s'_i$ and $s_i$ and $s'_i$ are compatible.
This notion of dominance, that we denote by \emph{NW}, was introduced in \cite{MS97}.

Finally, recall that two strategies $s_i$ and $s'_i$ of player $i$ are called
\oldbfe{payoff equivalent} if
\[
p_j(s_i, s_{-i}) = p_j(s'_i, s_{-i})
\]
for all $j \in [1..n]$ and all $s_{-i} \in S_{-i}$.
We denote this binary relation on the strategies by {\it PE}. 

These notions have natural counterparts for mixed strategies
that will be introduced later in the paper.

Each binary dominance relation $R$, so in particular \emph{W, S, NW} or \emph{PE},
induces the following binary relation on strategic games
$G := (S_1, \LL, S_n, p_1, \LL, p_n)$ and $G' := (S'_1, \LL, S'_n, p_1, \LL, p_n)$:

\begin{quote}
$G \Ra_{\hspace{-1mm} R \: } G'$ iff $G \neq G'$ and for all $i \in [1..n]$ 
each $s_i \in S_i \setminus S'_i$ is $R$-dominated in $G$ by some $s'_i \in S'_i$.
\end{quote}
If all iterations of $\Ra_{\hspace{-1mm} R}$ starting in an initial game $G$ yield the same
final outcome, we say that $R$ is \oldbfe{order independent}.

\subsection{Background}

In the literature on dominance relations in finite strategic games 
several order independence results were established, to wit:

\begin{itemize}

\item \cite{GKZ90} and \cite{Ste90} proved it
for strict dominance by pure strategies,

\item \cite{Bor90,Bor93}
established it for his notion of (unary) dominance,

\item \cite{OR94} proved it
for strict dominance by mixed strategies,

\item \cite{MS97,MS00}) proved it for nice weak dominance
up to the addition or removal of the payoff equivalent strategies
and a renaming of strategies.

This implies the same form of order independence for weak dominance 
by pure strategies for the games $(S_1, \LL, S_n, p_1, \LL, p_n)$ satisfying
the following \oldbfe{transference of decisionmaker indifference
  (TDI)} condition:

\[
  \begin{array}{l}
\mbox{for all $i, j \in [1..n]$, $r_i, t_i \in S_i$ and $s_{-i} \in S_{-i}$} \\
\mbox{$p_{i}(r_i, s_{-i}) = p_{i}(t_i, s_{-i})$ implies $p_{j}(r_i, s_{-i}) = p_{j}(t_i, s_{-i})$.}
  \end{array}
\]
Informally, this condition states that whenever for player $i$ two of
its strategies $r_i$ and $t_i$ are indifferent w.r.t.~some joint strategy
$s_{-i}$ of the other players, then $r_i$ and $t_i$ are also indifferent
w.r.t.~$s_{-i}$ for all players.

They also established analogous results for nice weak dominance and weak dominance
by mixed strategies.
\end{itemize}

These results were established by different methods and techniques.
In particular, the proof of order independence given in \cite{Bor90}
proceeds through a connection between the rationalizability notion of
\cite{Pea84} and the survival of a strategy under the iterated
dominance. In turn, the original proof of order independence for
strict dominance by mixed strategies given in \cite[ pages 61-62]{OR94}
involves in an analogous way a modification of the rationalizability
notion and relies on the existence of Nash equilibrium for strictly
competitive games.

It is useful to point out that the assumption that the games are finite
is crucial. In fact, in an interesting paper \cite{DS02} showed that in case of
infinite games order independence for strict dominance does not
hold. They also provided natural conditions under which the unique
outcome is guaranteed.

\subsection{Motivation}

In this paper we provide uniform and elementary proofs of the
abovementioned and related order independence results.
The table in Figure \ref{fig:classification} should clarify the scope of the paper.
So we deal both with unary and binary dominance relations and with pure and mixed
strategies. While binary dominance relations, such as the ones introduced in
the previous subsection, are more known, the unary ones,
introduced in \cite{Bor90,Bor93}, allow us to characterize a specific form of
the rational strategies.

\begin{figure}[htbp]
  \begin{center}
\begin{tabular}{|l|l|l|}
\hline
dominance \verb+\+ strategies & pure & mixed \\
\hline \hline
unary    &          & \\ \hline
binary     &          & \\
\hline
\end{tabular}
    \caption{Classification of the order independence results}
    \label{fig:classification}
  \end{center}
\end{figure}

Further, we also consider combinations of binary dominance relations, both for
pure and for mixed strategies.

Having in mind such a plethora of possibilities it is difficult to expect a single
`master result' that would imply all the discussed order independence results.
Still, as we show, it is possible to provide uniform proofs of these results
based on the same principles.
Notably, our presentation focuses on so-called abstract reduction
systems (see, e.g., \cite{Ter03}) in particular on Newman's Lemma
(see \cite{New42}) and some of its natural refinements.

Newman's Lemma offers a simple but highly effective and versatile tool for
proving order independence results.  We discuss it and its
consequences in detail in the next section and later, in Section \ref{ars-more}.
Let us just mention here
that it deals with the properties of a binary relation $\myra$ on an
arbitrary set $A$. Below $\tra$ denotes the
transitive reflexive closure of $\myra$.  We say that $\myra$ is \oldbfe{weakly
  confluent} if for all $a,b,c \in A$

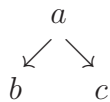
\begin{center}
$a$                                             \\
$\swarrow$ $\searrow$                           \\
$b$\ \ \ \ \ \ \ $c$                            \\
\end{center}

\NI
implies that for some $d \in A$

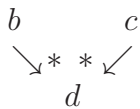
\begin{center}
$b$\ \ \ \ \ \ \ \ \ \ $c$                        \\
$\searrow \!*$ \ $\!*\swarrow$                    \\
$d$

\end{center}
Then Newman's lemma simply states that whenever

\begin{itemize}
\item no infinite $\myra$ sequences exist,

\item $\myra$ is weakly confluent,

\end{itemize}
then for each element $a \in A$ all $\myra$ sequences starting in $a$
have a unique `end outcome'.

It turns out that to prove order independence of a (binary or
unary) dominance relation $R$ it suffices to establish weak confluence
of the corresponding reduction relation $\Ra_{\hspace{-1mm} R}$ and
apply Newman's lemma.  In fact, since only finite games are
considered, no infinite $\Ra_{\hspace{-1mm} R}$ sequences exist.

To deal with combinations of two dominance relations, in particular 
the combination of nice weak dominance \emph{NW}
and payoff equivalence \emph{PE}, a relativized version of Newman's
lemma is helpful, where one only claims unique `end outcome' up to an
equivalence relation.  In the game-theoretic setting this equivalence relation
is an `equivalence up to strategy renaming' relation on strategic games.

Further, the following notion involving a relative dependence between two binary 
relations $\myra_1$ and $\myra_2$ on some set $A$ turns out to be useful.
We say that
$\myra_{\hspace{-1mm} 1}$ \oldbfe{left commutes with}
$\myra_{\hspace{-1mm} 2}$ if
\[
\myra_{\hspace{-1mm} 1 \: } \circ \myra_{\hspace{-1mm} 2 \: } \sse 
\myra_{\hspace{-1mm} 2 \: } \circ \myra^{\hspace{-1mm} *}_{\hspace{-1mm} 1 \: },
\]
i.e., if for all $a,b,c \in A$ \ 
$a \myra_{\hspace{-1mm} 1 \: } b \myra_{\hspace{-1mm} 2 \: } c$
implies that for some $d \in A$ \ 
$a \myra_{\hspace{-1mm} 2 \: } d \myra^{\hspace{-1mm} *}_{\hspace{-1mm} 1 \: } c$.

Now, one can prove that
$\Ra_{\hspace{-1mm} {\it PE} \: }$ left commutes with $\Ra_{\hspace{-1mm} {\it NW}}$.
This allows us to `push' the removal of the payoff equivalent strategies
to the `end' and prove a `structured' form of the order independence of \emph{NW}
combined with \emph{PE}, a result originally established in \cite{MS97}.


Our presentation is also influenced by \cite{GKZ90} where
order independence for strict dominance was proved by establishing
this result first for arbitrary (binary) dominance relations that are strict
partial orders and hereditary.

In our approach we isolate other useful properties of dominance
relations, both for the case of pure and mixed strategies.  In
particular, we identify conditions that allow us to conclude order
independence up to a renaming of strategies for a combination of two
reduction relations.  This allows us to identify the
relevant properties of nice weak dominance that lead to the results of
\cite{MS97}.

Of course, each strategy elimination procedure needs to be motivated,
either by clarifying the reasoning used by the players or by
clarifying its effect on the structure of the game, for example on 
its set of Nash equilibria. In our exposition we ignore these issues
since we focus on the dominance relations and the entailed
elimination procedures that were introduced and motivated
in the cited references.

\subsection{Organization of the paper}

The paper is organized as follows.  In the next section we discuss 
Newman's lemma.  Then in Section
\ref{sec:dominance}, following \cite{GKZ90}, we set the stage by
discussing (binary) dominance relations for strategic games and their
natural properties, in particular \emph{hereditarity}.  Intuitively, a
dominance relation is hereditary if it is inherited from a game to any
restriction.  Some dominance relations are hereditary, while others not.
Usually, for non-hereditary dominance relations order independence
does not hold.

Then, in Section \ref{sec:bor}, we generalize the approach of
\cite{Bor90,Bor93} to deal with arbitrary non-hereditary dominance
relations.  Informally, each such binary dominance relation $R$ can be modified
to a unary dominance relation for which under some natural assumption
both entailed reduction relations $\myra_{\hspace{-1mm} inh-R}$ and
$\Ra_{\hspace{-1mm} inh-R}$ are order independent.

Next, in Section \ref{sec:mixeddom} we study dominance relations where
the dominating strategies are mixed. We mimic here the development of
Section \ref{sec:dominance} by identifying natural properties and
establishing a general result on order independence.  We apply then
these results to show order independence for strict dominance by 
mixed strategies.
In Section \ref{sec:borm} we generalize the approach of Section \ref{sec:bor}
to the case when the dominating strategies are mixed.

To prepare the ground for results involving game equivalence we
discuss in Section \ref{ars-more} a modification of Newman's Lemma in
presence of an equivalence relation.  Then in Section
\ref{sec:pure-equ} we resume the discussion of dominance relations by
focusing on the payoff equivalence.  For this
dominance relation order independence does not hold, but order
independence up to a renaming of strategies does hold.  Analogous
results hold in case of equivalence to a mixed strategy
and are discussed in Section \ref{sec:randomized}.

Then in Section \ref{sec:combining} we study conditions under which
order dominance up to a renaming of strategies can be proved for a
combination of two dominance relations.  Such a combination is useful
to study when one of these two relations is not hereditary.  Then in
Section \ref{sec:nweak} we apply the obtained general result to get a
simple and informative proof of a result of \cite{MS97} that nice weak
dominance is order independent up to the removal of the payoff
equivalent strategies and a renaming of strategies.  In the
next two Sections, \ref{sec:combiningm} and \ref{sec:mnweak}, we mimic
these developments for the case of equivalence to and dominance by a
mixed strategy.  Finally, in the concluding section we summarize the
results in a tabular form and explain why each of the discussed order
independence results has to be established separately.

\section{Abstract Reduction Systems}
\label{sec:ars}

We provide first completely general results concerning abstract
reduction systems. An \oldbfe{abstract reduction system}, see, e.g.,
\cite{Hue80}, (and \cite{Ter03} for a more recent account, where a
slightly different terminology is used) is a pair $(A,\myra)$ where
$A$ is a set and $\myra$ is a binary relation (a \emph{reduction}) on
$A$.  Let $\myra^{\hspace{-1mm} +}$ denote the transitive closure of
$\myra$ and $\tra$ the transitive reflexive closure of $\myra$.  So in
particular, if $a = b$, then $a \tra b$.  Further, $a
\myra^{\hspace{-1mm} \epsilon} b$ means $a = b$ or $a \myra b$.

\begin{itemize}
  
\item We say that $b$ is a $\myra$-\oldbfe{normal form of} $a$ if $a
  \tra b$ and no $c$ exists such that $b \myra c$, and omit the
  reference to $\myra$ if it is clear from the context.  
  If every element of $A$ has a unique normal form, we say that
  $(A,\myra)$ (or just $\myra$ if $A$ is clear from the context)
  satisfies the \oldbfe{unique normal form property}.
\footnote{
We stress the fact that this notion of a normal form, standard in the theory
of abstract reduction systems, has no
    relation whatsoever to the notion of a game in normal form, another name used
    for strategic games.  In particular, the reader should bear in
    mind that later we shall consider \emph{strategic} games that are
    \emph{normal forms} of specific reduction relations on strategic games.
}

\item We say that $\myra$ is \oldbfe{weakly confluent}
if for all $a,b,c \in A$ 

\begin{center}
$a$                                             \\
$\swarrow$ $\searrow$                           \\
$b$\ \ \ \ \ \ \ $c$                            \\
\end{center}

\NI
implies that for some $d \in A$

\begin{center}
$b$\ \ \ \ \ \ \ \ \ \ $c$                        \\
$\searrow \!*$ \ $\!*\swarrow$                    \\
$d$

\end{center}

\item Following \cite{GKZ90} we say that
$\myra$ is \oldbfe{one step closed} if
for all $a \in A$ some $a' \in A$ exists such that
\begin{itemize}
\item $a \myra^{\hspace{-1mm} \epsilon} a'$,

\item if $a \myra b$, then $b \myra^{\hspace{-1mm} \epsilon} a'$.
\end{itemize}
\end{itemize}

In all proofs of weak confluence given in the paper we shall actually establish that for some $d \in A$
we have $b \myra^{\hspace{-1mm} \epsilon} d$ and $c \myra^{\hspace{-1mm} \epsilon} d$.

In the sequel, as already mentioned, we shall repeatedly rely upon
the following lemma established in \cite{New42}. 

\begin{lemma} [Newman] \label{lem:newman}
Consider an abstract reduction system $(A,\myra)$ such that

\begin{itemize}
\item no infinite $\myra$ sequences exist,

\item $\myra$ is weakly confluent.

\end{itemize}
Then $\myra$ satisfies the unique normal form property.
\end{lemma}

\Proof 
(Taken from \cite[ page 15]{Ter03}.)

\NI By the first assumption every element of $A$ has a normal
form.  To prove uniqueness, call an element $a$ \emph{ambiguous} if it
has at least two different normal forms.  We show that for
every ambiguous $a$ some ambiguous $b$ exists such that $a \myra b$.
This proves absence of ambiguous elements by the first assumption.

So suppose that some element $a$ has two distinct normal forms $n_1$
and $n_2$.  Then for some $b, c$ we have $a \myra b \tra n_1$ and $a
\myra c \tra n_2$.  By weak confluence some $d$ exists such that $b
\tra d$ and $c \tra d$. Let $n_3$ be a normal form of $d$. It is also
a normal form of $b$ and of $c$. Moreover $n_3 \neq n_1$ or $n_3 \neq
n_2$. If $n_3 \neq n_1$, then $b$ is ambiguous and $a \myra b$.  
And if $n_3 \neq n_2$, then $c$ is ambiguous and $a \myra c$.  
\HB
\VV

Note that if $\myra$ is not irreflexive, then the first condition
is violated. So this lemma can be applicable only to the relations
$\myra$ that are irreflexive.  All reduction relations
on games here considered are by definition irreflexive.
Moreover, because the games are assumed to be finite, these
reduction relations automatically satisfy the first condition 
of Newman's lemma.

Also, the following simple observation will be helpful.

\begin{note} [Unique Normal Form] \label{lem:uni}
Consider two abstract reduction systems $(A, \myra_{\hspace{-1mm}
  1 \: })$ and $(A, \myra_{\hspace{-1mm} 2 \: })$ such that

\begin{itemize}
\item $\myra_{\hspace{-1mm} 1 \: }$ satisfies the unique normal form property,

\item $\myra_{\hspace{-1mm} 1}^{\hspace{-1mm} +} = \myra_{\hspace{-1mm} 2}^{\hspace{-1mm} +}$.

\end{itemize}
Then $\myra_{\hspace{-1mm} 2}$ satisfies the unique normal form property.
\HB
\end{note}

In the remainder of the paper we shall study abstract reduction
systems that consist of the set of all restrictions of a game and a
reduction relation on them. Since we limit ourselves to finite games,
in such abstract reduction systems $(A,\myra)$ no infinite $\myra$ sequences exist.

In this context there are three natural ways of establishing that $(A,\myra)$
satisfies the unique normal form property:

\begin{itemize}
\item by showing that $\myra$ is one step closed: this directly implies weak
confluence, and then Newman's Lemma can be applied;

\item by showing that $\myra$ is weakly confluent and applying 
Newman's Lemma;

\item by finding a `more elementary' reduction relation $\myra_{\hspace{-1mm} 1}$
such that 

\begin{itemize}

\item no infinite $\myra_{\hspace{-1mm} 1}$ sequences exist,

\item $\myra_{\hspace{-1mm} 1}$ is weakly confluent,

\item $\myra_{\hspace{-1mm} 1}^{\hspace{-1mm} +} = \myra^{\hspace{-1mm} +}$,

\end{itemize}
and applying Newman's Lemma and the Unique Normal Form Note \ref{lem:uni}.

\end{itemize}
For some reduction relations all three results are equally easy to
establish, while for some others only one.

\section{Dominance Relations}
\label{sec:dominance}

We now study (binary) dominance relations in full generality.
A \oldbfe{dominance relation} is a function that assigns to each
game $G := (S_1, \LL, S_n, p_1, \LL, p_n)$ a subset $R_G$
of
$
\bigcup_{i = 1}^{n} (S_i \times S_i).
$
Instead of writing that $s_i \ R_G \ s'_i$ holds we write that
$s_i \ R \ s'_i$ \oldbfe{holds for} $G$. We say then 
that \oldbfe{$s_i$ is
$R$-dominated by $s'_i$ in $G$} or that
that \oldbfe{$s'_i$ $R$-dominates $s_i$ in $G$}.
When $G$ is clear from the context we drop a reference to it and view a dominance
relation as a binary relation on the strategies of $G$.

Given a dominance relation $R$ 
we introduce two notions of reduction between a game 
$G := (S_1, \LL, S_n, p_1, \LL, p_n)$ and its restriction $G' := (S'_1, \LL, S'_n,$ $p_1, \LL, p_n)$.

\begin{itemize}
\item We write $G \myra_{\hspace{-1mm} R \: } G'$ when $G \neq G'$ and for all $i \in [1..n]$ 

\[
\mbox{each $s_i \in S_i \setminus S'_i$ is $R$-dominated in $G$ by some $s'_i \in S_i$}.
\]

\item We write $G \Ra_{\hspace{-1mm} R \: } G'$ when $G \neq G'$ and for all $i \in [1..n]$ 

\[
\mbox{each $s_i \in S_i \setminus S'_i$ is $R$-dominated in $G$ by some $s'_i \in S'_i$}.
\]
\end{itemize}

So the relations $\myra_{\hspace{-1mm} R}$ and $\Ra_{\hspace{-1mm} R}$
differ in just one symbol (spot the difference).  Namely, in the case
of $\myra_{\hspace{-1mm} R \: }$ we require that each strategy removed
from $S_i$ is $R$-dominated in $G$ by a strategy in $S_i$, while in
case of $\Ra_{\hspace{-1mm} R \: }$ we require that each strategy
removed from $S_i$ is $R$-dominated in $G$ by a strategy in $S'_i$. So
in the latter case the dominating strategy should not be removed at
the same time.  

In the literature both reduction relations were considered.
In our subsequent presentation we shall focus on the second one, $\Ra_{\hspace{-1mm} R}$,
since

\begin{itemize}
\item for most of the reduction relations studied here 
$\myra_{\hspace{-1mm} R}$ and $\Ra_{\hspace{-1mm} R}$ coincide,

\item for payoff equivalence these relations do not coincide
and only the second reduction relation is meaningful.
\end{itemize}

On the other hand, the first reduction
relation, $\myra_{\hspace{-1mm} R}$, allows us to define the `maximal'
elimination strategy according to which in each round all
$R$-dominated strategies are deleted.  Such a natural strategy is 
in particular of interest
when order independence fails, see, e.g., \cite{Gil02}.  

Further, note when $G \myra_{\hspace{-1mm} R \: } G'$, the game $G'$
can be `degenerated' in the sense that some of the strategy sets of
$G'$ can be empty. However, this cannot happen when
$\myra_{\hspace{-1mm} R}$ and $\Ra_{\hspace{-1mm} R}$ coincide, since
then $G \Ra_{\hspace{-1mm} R \: } G'$ implies that $G'$ is not `degenerated'.

Finally, let us mention that for various type of dominance relations
$R$ (unary or binary, for pure and mixed strategies) studied here the
equivalence between the corresponding $\myra_{\hspace{-1mm} R}$ and
$\Ra_{\hspace{-1mm} R}$ reduction relations plays a crucial role in
the proofs of the order independence results.

So each reduction relation has some advantages and it is natural to introduce both of them.

Recall that a strict partial order is an irreflexive transitive
relation.  We say now that a dominance relation $R$ is a
\oldbfe{strict partial order} if for each game $G$ the binary relation
$R_G$ is a strict partial order and reuse in a similar way other
typical properties of binary relations.  The following observation
clarifies the first item above and will be needed later.

\begin{lemma}[Equivalence] \label{lem:equ1}
If a dominance relation $R$ is a strict partial order, then
the relations $\myra_{\hspace{-1mm} R}$  and $\Ra_{\hspace{-1mm} R}$ coincide.
\end{lemma}
\Proof It suffices to show that if $G \myra_{\hspace{-1mm} R \: } G'$, then $G \Ra_{\hspace{-1mm} R \: } G'$.

Let $G := (S_1, \LL, S_n, p_1, \LL, p_n)$ and $G' := (S'_1, \LL, S'_n,
p_1, \LL, p_n)$.  Suppose that some $s_i \in S_i \setminus S'_i$ is
$R$-dominated in $G$ by some $s'_i \in S_i$.  
Since $R$ is a strict partial order and $S_i$ is finite,
a strategy $s'_i \in S_i$ exists
that $R$-dominates $s_i$ in $G$ and is not $R$-dominated
in $G$.  So this $s'_i$ is not eliminated in the step $G \myra_{\hspace{-1mm} R \: } G'$ and
consequently $s_i$ is $R$-dominated in $G$ by some $s'_i \in S'_i$.
\HB 
\VV

In what follows we establish a general `order independence' result for
the reduction relation $\Ra_{\hspace{-1mm} R}$ for the dominance
relations $R$ that are strict partial orders and satisfy the following
natural assumption due to \cite{GKZ90}. We say that a dominance
relation $R$ is \oldbfe{hereditary} if for every game $G$, its restriction
$G'$ and two strategies $s_i$ and $s'_i$ of $G'$
\[
\mbox{$s_i$ is $R$-dominated by $s'_i$ in $G$ implies $s_i$ is $R$-dominated by $s'_i$ in $G'$.}
\]

Each reduction relation $\Ra_{\hspace{-1mm} R}$ can be specialized by
stipulating that a \emph{single strategy} is removed.  We denote the
corresponding reduction relation by $\Ra_{\hspace{-1mm} 1, R}$.  A
natural question when the reduction relation $\Ra_{\hspace{-1mm} R}$
can be modeled using the iterated application of the
$\Ra_{\hspace{-1mm} 1,R}$ reduction relation does
not turn out to be interesting.

In fact, for most dominance relations that are of importance such a
modeling is not possible.  The reason is that when removing strategies
in an iterated fashion, in particular in the one-at-a-time fashion,
some previously undominated strategies can become eligible for
removal.  So this process can yield a different outcome than a single
removal of several strategies.

In contrast, the following definition seems to capture a relevant
property.  We say that that a reduction relation 
$\Ra_{\hspace{-1mm} R}$ satisfies the
\oldbfe{one-at-a-time} property if 
\[
\Ra^{\hspace{-1mm} +}_{\hspace{-1mm} 1,R} = \Ra^{\hspace{-1mm} +}_{\hspace{-1mm} R}. 
\]
Obviously, if
$\Ra^{\hspace{-1mm} +}_{\hspace{-1mm} 1,R} = \Ra^{\hspace{-1mm}
  +}_{\hspace{-1mm} R}$, then also $\Ra^{\hspace{-1mm}
  *}_{\hspace{-1mm} 1,R} = \Ra^{\hspace{-1mm} *}_{\hspace{-1mm} R}$.
The following result clarifies when the
one-at-a-time property holds.

\begin{theorem}[One-at-a-time Elimination] \label{thm:one}
  For a dominance relation $R$ that is hereditary the $\Ra_{\hspace{-1mm}
    R}$ relation satisfies the one-at-a-time property.
\end{theorem}

\Proof
Note that always $\Ra_{\hspace{-1mm} 1,R} \sse \Ra_{\hspace{-1mm} R}$, so
$
\Ra^{\hspace{-1mm} +}_{\hspace{-1mm} 1, R} \sse \Ra^{\hspace{-1mm} +}_{\hspace{-1mm} R}
$
always holds.

To prove the inverse inclusion it suffices to show that
$\Ra_{\hspace{-1mm} R} \sse \Ra^{\hspace{-1mm} +}_{\hspace{-1mm} 1, R}.
$
So suppose that $G \Ra_{\hspace{-1mm} R \: } G'$.
We prove that $G \Ra^{\hspace{-1mm} +}_{\hspace{-1mm} 1, R \: } G'$ by
induction on the number $k$ of strategies deleted in the transition 
from $G$ to $G'$.
If $k=1$, then $G \Ra_{\hspace{-1mm} 1, R \: } G'$ holds.

Suppose now that claim holds for some $k > 1$. Assume that
$
G := (S_1, \LL, S_n,$ $p_1, \LL, p_n)
$
and
$
G' := (S'_1, \LL, S'_n, p_1, \LL, p_n).
$
For each $i \in [1..n]$ let $S_i \setminus S'_i := \C{t^{1}_i, \LL, t^{k_i}_{i}}$.
So for all $i \in [1..n]$ and all $j \in [1..k_i]$ 
the strategy $t^{j}_{i}$ is $R$-dominated
in $G$ by some $s^{j}_{i} \in S'_i$. Choose some strategy $t^{j_0}_{i_0}$ and let
$G''$ be the game resulting from $G$ by removing $t^{j_0}_{i_0}$ from $S_{i_0}$.
Then $G \Ra_{\hspace{-1mm} 1, R \: } G''$.

Since $t^{j_0}_{i_0} \not\in \cup_{i=1}^{n} S'_{i}$ each strategy
$s^{j}_{i}$ is in $G''$.  So by the hereditarity of $R$ each strategy
$t^{j}_i$, where $(i,j) \neq (i_0, j_0)$, is $R$-dominated in $G''$ by
$s^{j}_{i}$.  This means that $G'' \Ra_{\hspace{-1mm} R \: } G'$. By
the induction hypothesis $G'' \Ra^{\hspace{-1mm} +}_{\hspace{-1mm} 1,
  R \: } G'$, hence $G \Ra^{\hspace{-1mm} +}_{\hspace{-1mm} 1, R \: }
G'$.  
\HB 
\VV

Now, given a dominance relation $R$ that is hereditary and is a strict
partial order we can establish that the
$\Ra_{\hspace{-1mm} R}$ reduction
relation on the set of all restrictions
of a game $H$ satisfies the unique normal form property
(in short: \oldbfe{is UN}) in one of the following three ways:

\begin{itemize}
\item by showing that $\Ra_{\hspace{-1mm} R}$ is one step closed;
this is the argument provided by \cite{GKZ90},

\item by proving that $\Ra_{\hspace{-1mm} R}$ 
is weakly confluent,

\item by proving that $\Ra_{\hspace{-1mm} 1, R}$ 
is weakly confluent. 
\end{itemize}

In the last case one actually proceeds by showing that 
$\Ra_{\hspace{-1mm} 1, R}$ 
satisfies the diamond property, where 
we say that $\myra$ 
satisfies the 
\oldbfe{diamond property} if
for all $a,b,c \in A$ such that $b \neq c$

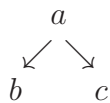
\begin{center}
$a$                                             \\
$\swarrow$ $\searrow$                           \\
$b$\ \ \ \ \ \ \ $c$                            \\
\end{center}

\NI
implies that for some $d \in A$

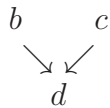
\begin{center}
$b$\ \ \ \ \ \ \ $c$                            \\
$\searrow$ $\swarrow$                           \\
$d$

\end{center}

All three proofs are straightforward.
As an illustration we provide the proof for the second approach
as its pattern will be repeated a number of times.

\begin{lemma}[Weak Confluence] \label{lem:diamond1}
Consider a dominance relation $R$ that is hereditary and is a strict partial order.
Then the  $\Ra_{\hspace{-1mm} R}$ relation on the set of all restrictions
of a game $H$ is weakly confluent.
\end{lemma}

\Proof
Suppose
\begin{center}
\ $G$            \\[-4mm]
\vspace{2mm}
$\Swarrow$ \ \ \ \ \ \ \ \\[-4mm]
$\hspace{3mm}_{R}$ \ $_{R}\hspace{-1mm}\Searrow$                           \\
\ \ $G'$\ \  \ \ \ $G''$           
\end{center}

\NI
We prove that then
\begin{center}
\ \ $G'$\ \  \ \ \ \ \ \ $G''$       \\[-4mm]
\vspace{2mm}
\ \ \ \ \ \ \ \ \ $\Swarrow$  \\[-4mm]
\ $_{R} \hspace{-1mm}\Searrow\! ^{\epsilon}$ \ \ $^{\epsilon}\! \hspace{2mm}_{R}$                           \\
$\ \ G' \cap G''$
\end{center}

\NI
Recall that $a \Ra^{\hspace{-1mm} \epsilon}_{\hspace{-1mm} R \: } b$ means $a \Ra_{\hspace{-1mm} R \: } b$ or $a = b$.

If  $G'$ is a restriction of $G' \cap G''$, then 
$G' = G' \cap G''$ and consequently $G' \Ra^{\hspace{-1mm} \epsilon}_{\hspace{-1mm} R \: } G' \cap G''$.
Otherwise suppose
\[
G' := (S'_1, \LL, S'_n, p_1, \LL, p_n),
\]
\[
G'' := (S''_1, \LL, S''_n, p_1, \LL, p_n).
\]
Then
\[
G' \cap G'' = (S'_1 \cap S''_1, \LL, S'_n \cap S''_n, p_1, \LL, p_n).
\]
Fix $i \in [1..n]$ and consider a strategy $s_i \in
S'_i$ such that $s_i \not\in S'_i \cap S''_i$.  So $s_i$ is eliminated in
the step $G \Ra_{\hspace{-1mm} R \: } G''$.  Hence some $s'_i \in S_i$ $R$-dominates
$s_i$ in $G$.  
\II

\NI
\emph{Case 1.} $s'_i \in S'_i$. 

$G'$ is a restriction of $G$ and $R$ is hereditary so
$s'_i$ also $R$-dominates $s_i$ in $G'$.
\II

\NI
\emph{Case 2.} $s'_i \not \in S'_i$. 

So $s'_i$ is eliminated in the step $G \Ra_{\hspace{-1mm} R \: } G'$.  Hence 
a strategy $s''_i \in S'_i$ exists
that $R$-dominates $s'_i$ in $G$.  By the transitivity of $R$, $s''_i$
$R$-dominates $s_i$ in $G$ and hence, by hereditarity, in $G'$.

\III

This proves $G' \myra_{\hspace{-1mm} R \: } G' \cap G''$
and hence, by the
Equivalence Lemma \ref{lem:equ1},  
$G' \Ra_{\hspace{-1mm} R \: } G' \cap G''$.

By symmetry $G'' \Ra^{\hspace{-1mm} \epsilon}_{\hspace{-1mm} R \: }  G' \cap G''$. 
\HB
\VV

This brings us to the following result of \cite{GKZ90}.

\begin{theorem}[Elimination] \label{thm:hereditary}
For a dominance relation $R$ that is hereditary and a strict partial order
the  $\Ra_{\hspace{-1mm} R \: }$ relation is UN.
\HB
\end{theorem}

To illustrate a direct use of the above results consider the strict
dominance relation $S$.
It entails the reduction relation
$\Ra_{\hspace{-1mm} S}$ on games obtained by instantiating $R$ in
$\Ra_{\hspace{-1mm} R}$ by the strict dominance relation.  As already
noted by \cite{GKZ90} strict dominance is clearly hereditary and is a
strict partial order. So we get the following conclusion.

\begin{theorem}[Strict Elimination] \label{thm:strict1}

\mbox{} \vspace{-3mm} 
  \begin{enumerate} \smallromani

\item 
The  $\Ra_{\hspace{-1mm} S}$ relation is UN.

\item 
The  $\Ra_{\hspace{-1mm} S}$ relation satisfies the one-at-a-time property.
\HB
\end{enumerate}
\end{theorem}

In other words, the process of iterated elimination of strictly
dominated strategies yields a unique outcome and coincides with the
outcome of the iterated elimination of a single dominated strategy.

\section{Pure Strategies: Inherent Dominance}
\label{sec:bor}

In this section we introduce and study a natural generalization of the
binary dominance notion, due to \cite{Bor90,Bor93}.  Consider a game $(S_1, \LL,
S_n,$ $p_1, \LL, p_n)$.  Let $R$ be a dominance relation and
$\tilde{S}_{-i}$ a non-empty subset of
$S_{-i}$.  We say that a strategy $s_i$ is $R$-\oldbfe{dominated
  given $\tilde{S}_{-i}$} by a strategy $s'_i$ if $s_i$ is
$R$-dominated by $s'_i$ in the game $(S_i, \tilde{S}_{-i}, p_1, \LL,
p_n)$.  Then we say that a strategy $s_i$ is \oldbfe{inherently
  $R$-dominated} if for every non-empty subset $\tilde{S}_{-i}$ of
$S_{-i}$ it is $R$-dominated given $\tilde{S}_{-i}$ by some strategy
$s'_i$. So we turned in this way the binary relation $R$ to a
unary relation on the strategies.

Note that in the definition of inherent $R$-dominance for each subset
$\tilde{S}_{-i}$ of $S_{-i}$ a different strategy of player $i$ can
$R$-dominate the considered strategy $s_i$. This can make this notion
of dominance stronger than $R$-dominance.  \cite{Bor90,Bor93} studied
this notion of dominance for $R$ being weak dominance
and established for it the order independence.
The resulting dominance relation, inherent weak dominance, is an intermediate
notion between strict and weak dominance. Indeed, 
it is clearly implied by strict dominance and implies in turn weak
dominance. The converse implications do not hold
as the following two examples show.
In the game
\begin{center}
\begin{game}{3}{2}
      & $L$    & $R$ \\
$T$   &$2,-$   &$1,-$\\
$M$   &$1,-$   &$2,-$ \\
$B$   &$1,-$   &$3,-$
\end{game}
\end{center}
the strategy $M$ is weakly dominated by $T$ given $\C{L}$
and weakly dominated by $B$ given $\C{R}$ or given $\C{L,R}$.
So $M$ is inherently weakly dominated but is not strictly dominated
by any strategy.

In turn in the game

\begin{center}
\begin{game}{2}{2}
      & $L$    & $R$\\
$T$   &$2,-$   &$1,-$\\
$B$   &$1,-$   &$1,-$
\end{game}
\end{center}
the strategy $B$ is not inherently weakly dominated but is
weakly dominated.

It is well-known that weak dominance is not order independent.
We shall return to this matter in Section \ref{sec:nweak}.
The intuitive reason is that weak dominance is not hereditary.
As a consequence the proof of
the corresponding weak confluence property does not go through.  

The notion of inherent $R$-dominance does not fit into the framework
developed in Section \ref{sec:dominance}, since it is a unary relation.
However, when studying reduction by means of it we
can proceed in a largely analogous fashion.
So first we introduce two notions of
reduction between a game $G := (S_1, \LL, S_n, p_1, \LL, p_n)$ and its
restriction $G' := (S'_1, \LL, S'_n, p_1, \LL, p_n)$, this time involving
the inherent $R$-dominance notion.

\begin{itemize}
\item We write $G \myra_{\hspace{-1mm} inh-R \: } G'$ when $G \neq G'$ and for all $i \in [1..n]$ 
\[
\mbox{each $s_i \in S_i \setminus S'_i$ is inherently $R$-dominated in $G$}.
\]

\item We write $G \Ra_{\hspace{-1mm} inh-R \: } G'$ when $G \neq G'$ and for all $i \in [1..n]$ 
for every non-empty subset $\tilde{S}_{-i}$ of $S_{-i}$
\[
\mbox{each $s_i \in S_i \setminus S'_i$ is $R$-dominated in $G$ given $\tilde{S}_{-i}$
by some $s'_i \in S'_i$}.
\]
\end{itemize}

So in the $\myra_{\hspace{-1mm} inh-R \: }$ relation for every non-empty
subset $\tilde{S}_{-i}$ of $S_{-i}$ we require $R$-dominance in $G$
given $\tilde{S}_{-i}$ by some $s'_i \in S_i$, while in the
$\Ra_{\hspace{-1mm} inh-R \: }$ relation for every non-empty subset
$\tilde{S}_{-i}$ of $S_{-i}$ we require $R$-dominance in $G$ given
$\tilde{S}_{-i}$ by some $s'_i \in S'_i$.  \cite{Bor90,Bor93} considered the
first relation, $\myra_{\hspace{-1mm} inh-R \: }$, for $R$ being weak dominance.
We introduce the
second one, $\Ra_{\hspace{-1mm} inh-R \: }$, to streamline the
presentation. As in Section \ref{sec:dominance} under a
natural assumption both notions turn out to be
equivalent.

\begin{lemma}[Equivalence] \label{lem:equ3} 
For a dominance relation $R$ that is a strict partial order
the relations $\myra_{\hspace{-1mm} inh-R}$  and $\Ra_{\hspace{-1mm} inh-R}$ coincide.
\end{lemma}

\Proof The proof is similar to that of the Equivalence Lemma
\ref{lem:equ1}.  It suffices to show that if $G \myra_{\hspace{-1mm}
  inh-R \: } G'$, then $G \Ra_{\hspace{-1mm} inh-R \: } G'$.

Let $G := (S_1, \LL, S_n, p_1, \LL, p_n)$ and $G' := (S'_1, \LL, S'_n,
p_1, \LL, p_n)$.  Suppose that some $s_i \in S_i \setminus S'_i$ is
inherently $R$-dominated in $G$. Let $\tilde{S}_{-i}$ be a non-empty subset of
$S_{-i}$.  Some strategy $s'_i \in S_i$ \ $R$-dominates $s_i$
in $G$ given $\tilde{S}_{-i}$.  $R$ is a strict partial order
and $S_i$ is finite, so a
strategy $s'_i \in S_i$ exists that $R$-dominates $s_i$ in $G$
given $\tilde{S}_{-i}$ and is not $R$-dominated in $G$ given
$\tilde{S}_{-i}$ by any strategy in $S_i$.  
So this $s'_i$ is not eliminated in the step $G
\myra_{\hspace{-1mm} inh-R \: } G'$ and consequently $s_i$ is 
$R$-dominated in $G$ given $\tilde{S}_{-i}$ by some $s'_i \in S'_i$.
\HB 
\VV

The following simple observation relates the 
$\Ra_{\hspace{-1mm} inh-R}$ reduction relation to the
previously introduced relation $\Ra_{\hspace{-1mm} R}$.

\begin{note}[Comparison] \label{not:comparison}
Consider a dominance relation $R$. Then
  \begin{enumerate} \smallromani

\item $\Ra_{\hspace{-1mm} inh-R} \sse \Ra_{\hspace{-1mm} R}$.

\item If $R$ is hereditary, then
the relations $\Ra_{\hspace{-1mm} inh-R}$ and $\Ra_{\hspace{-1mm} R}$ coincide.
\HB
\end{enumerate}
\end{note}

So for hereditary dominance relations no new reduction relations
were introduced here. Further, it is easy to provide examples of a
non-hereditary $R$, for instance weak dominance, for which the
reduction relations $\Ra_{\hspace{-1mm} inh-R}$ and
$\Ra_{\hspace{-1mm} R}$ differ.

We now establish order independence for specific
$\Ra_{\hspace{-1mm} inh-R}$ reduction relations.
Following \cite{GKZ90} we say that a dominance relation $R$ satisfies
the \oldbfe{individual independence of irrelevant alternatives} condition (in
short, \oldbfe{IIIA}) if for every game $(S_i, S_{-i}, p_1, \LL, p_n)$ 
the following holds:
\[
  \begin{array}{l}
\mbox{for all $i \in [1..n]$, all non-empty $S'_i \sse S_i$ and $s_i, s'_i \in S'_i$} \\
\mbox{$s_i \ R \ s'_i$ holds in $(S_i, S_{-i}, p_1, \LL, p_n)$ iff it
 holds in $(S'_i, S_{-i}, p_1, \LL, p_n)$.}
  \end{array}
\]

IIIA is a very reasonable condition. All specific dominance relations
considered in this paper satisfy it.

\begin{lemma}[Weak Confluence] \label{lem:diamond3}
For a dominance relation $R$ that satisfies the IIIA condition and is a strict partial order
the  $\Ra_{\hspace{-1mm} inh-R}$ relation on the set of all restrictions
of a game $H$ is weakly confluent.
\end{lemma}

\Proof We proceed as in the proof of
the Weak Confluence Lemma \ref{lem:diamond1}.
Suppose $G \Ra_{\hspace{-1mm} inh-R \: } G'$ and $G \Ra_{\hspace{-1mm}
  inh-R \: } G''$.  We prove that then $G' \Ra^{\epsilon}_{\hspace{-1mm}
 \:  inh-R \: } G' \cap G''$ and $G'' \Ra^{\epsilon}_{\hspace{-1mm}  \: inh-R \: }
G' \cap G''$.

If $G'$ is a restriction of $G' \cap G''$, then $G' = G' \cap G''$ and
consequently $G' \Ra^{\epsilon}_{\hspace{-1mm}  inh-R \: } G' \cap G''$.
Otherwise suppose
$
G' := (S'_1, \LL, S'_n, p_1, \LL, p_n),
$
$
G'' := (S''_1, \LL, S''_n, p_1, \LL, p_n).
$
Then
$
G' \cap G'' = (S'_1 \cap S''_1, \LL, S'_n \cap S''_n, p_1, \LL, p_n).
$

Fix $i \in [1..n]$.  Consider a strategy $s_i \in
S'_i$ such that $s_i \not \in S'_i \cap S''_i$.  
So $s_i$ is eliminated in the step $G \Ra_{\hspace{-1mm}  \: inh-R \: }
G''$.
Take now a non-empty subset $\tilde{S}_{-i}$ of $S'_{-i}$
(and hence of $S_{-i}$). The strategy
$s_i$ is $R$-dominated given $\tilde{S}_{-i}$ in $G$ by
some strategy $s'_i \in S_i$.
\II

\NI
\emph{Case 1.} $s'_i \in S'_i$. 

Then, since $R$ satisfies the IIIA condition, $s_i \ R \ s'_i$ holds in the game
$(S'_i, \tilde{S}_{-i}, p_1, \LL, p_n)$, i.e., 
$s_i$ is $R$-dominated given $\tilde{S}_{-i}$ in $G'$.

\II

\NI
\emph{Case 2.} $s'_i \not \in S'_i$. 

So $s'_i$ is eliminated in the step $G \Ra_{\hspace{-1mm}  \: inh-R \: }
G'$.  Hence a strategy $s''_i
\in S'_i$ exists that $R$-dominates $s'_i$ in $G$ given
$\tilde{S}_{-i}$.  By the transitivity of $R$ the strategy $s''_i$ \ $R$-dominates $s_i$ in
$G$ given $\tilde{S}_{-i}$ and hence, since $R$ satisfies the IIIA condition,
$s_i$ is $R$-dominated given $\tilde{S}_{-i}$ in $G'$.
\III

So we showed that each strategy $s_i$ of player $i$ eliminated in the transition
from $G'$ to $G' \cap G''$ is inherently $R$-dominated in $G'$.
This proves $G' \myra_{\hspace{-1mm} inh-R \: } G' \cap G''$
and hence, by the
Equivalence Lemma \ref{lem:equ3}  
$G' \Ra_{\hspace{-1mm} inh-R \: } G' \cap G''$.

By symmetry $G'' \Ra^{\epsilon}_{\hspace{-1mm} \: inh-R \: } G' \cap G''$.  
\HB
\VV

We can now draw the desired conclusion using Newman's Lemma 
\ref{lem:newman}.

\begin{theorem}[Inherent Elimination] \label{thm:strict3}
For a dominance relation $R$ that satisfies the IIIA condition and is a strict partial order
the $\Ra_{\hspace{-1mm} \: inh-R \: }$ relation is UN.
\HB
\end{theorem}

As in Section \ref{sec:dominance} we introduce the 
$\Ra_{\hspace{-1mm} 1, inh-R}$ reduction relation that
removes exactly one strategy, and as before  we say that $\Ra_{\hspace{-1mm}
   inh-R}$ satisfies the \oldbfe{one-at-a-time} property when
\[
\Ra^{\hspace{-1mm} +}_{\hspace{-1mm} 1,inh-R} = \Ra^{\hspace{-1mm} +}_{\hspace{-1mm} inh-R}. 
\]

The following counterpart of the One-at-a-time
Elimination Theorem \ref{thm:one} then holds.

\begin{theorem}[One-at-a-time Elimination] \label{thm:oneB}
  For a dominance relation $R$ that satisfies the IIIA condition the relation $\Ra_{\hspace{-1mm}
   inh-R}$ satisfies the one-at-a-time property.
\end{theorem}

\Proof
Analogous to the proof of the One-at-a-time Elimination Theorem \ref{thm:one} and omitted.
\HB
\VV

Since the weak dominance relation $W$ satisfies the IIIA condition and
is a strict partial order, by the above results we get the following
counterpart of the Strict Elimination Theorem \ref{thm:strict1}.

\begin{theorem}[Inherent Weak Elimination] \label{thm:weak-inh}
\mbox{} \vspace{-3mm} 
  \begin{enumerate} \smallromani

\item 
The $\Ra_{\hspace{-1mm} inh-W}$ relation is UN.

\item 
The  $\Ra_{\hspace{-1mm} inh-W}$ relation satisfies the one-at-a-time property.
\HB
\end{enumerate}
\end{theorem}

The first item was established in \cite{Bor90}.
In \cite{Bor93} it was shown that a strategy is inherently weakly dominated
iff it is not rational, in the sense that it is not a best response to
a belief formed over the pure strategies of other players when their payoff
functions are not known --- it is only assumed that their payoff functions
are compatible with their publicly known preferences.
So the $\Ra_{\hspace{-1mm} inh-W}$ relation allows
us to model iterated removal of strategies that are not rational
in this sense.

\section{Mixed Dominance Relations}
\label{sec:mixeddom}

The notion of dominance studied in Section \ref{sec:dominance}
involved two pure strategies. In this section we study the
dominance relations in which the dominating strategies are mixed
and develop the appropriate general results.

Let us recall first the definitions.  Given a set of strategies $S_i$
available to player $i$, by a \oldbfe{mixed strategy} we mean a
probability distribution over $S_i$ and denote this set of mixed
strategies by $M_i$.

Given a mixed strategy $m_i$ we
define
\[
support(m_i) := \{s_i \in S_i \mid m_i(s_i) > 0\}.
\]

Consider a game $(S_1, \LL, S_n, p_1, \LL, p_n)$.
Each payoff function $p_i$ is generalized to a function
\[
p_i: M_1 \times \LL \times M_n \myra \cal{R}
\]
by putting for a sequence $(m_1, \LL, m_n)$ of mixed strategies from
$M_1 \times \LL \times M_n$
\[
p_i(m_1, \LL, m_n) := \sum_{s \in S} m_1(s_1) \: \LL \: m_n(s_n) \: p_i(s).
\]

As usual, we identify a mixed strategy for player $i$ of a restriction
$G'$ of $G$ with a mixed strategy of $G$ by assigning the probability
0 to the strategies of player $i$ that are present in $G$ but not in
$G'$. Further, we can view a mixed strategy for player $i$ in $G$ as a
mixed strategy in $G'$ if its support is a subset of the set of all
strategies of player $i$ in $G'$.  Also, we can identify each pure
strategy $s_i$ with the mixed strategy that assigns to $s_i$ the
probability 1.

A \oldbfe{mixed dominance relation} is a function that assigns to each
game $G := (S_1, \LL, S_n, p_1, \LL, p_n)$ a subset $R_G$ of
$
\bigcup_{i = 1}^{n} (S_i \times M_i).
$
When $s_i \ R_G \ m'_i$ holds we say that $s_i \ R \ m'_i$ \oldbfe{holds for}
$G$ and also say that 
\oldbfe{$s_i$ is $R$-dominated by $m'_i$ in $G$}, or that
\oldbfe{$m'_i$ $R$-dominates $s_i$ in $G$}.

As in Section \ref{sec:dominance}
we introduce now two notions of reduction between a game $G :=
(S_1, \LL, S_n, p_1, \LL, p_n)$ and its restriction $G' := (S'_1, \LL,
S'_n, p_1, \LL, p_n)$, this time involving a mixed dominance relation $R$.

\begin{itemize}
\item We write $G \myra_{\hspace{-1mm} R \: } G'$ when $G \neq G'$ and for all $i \in [1..n]$ 

\[
\mbox{each $s_i \in S_i \setminus S'_i$ is $R$-dominated in $G$ by some $m'_i \in M_i$}.
\]

\item We write $G \Ra_{\hspace{-1mm} R \: } G'$ when $G \neq G'$ and for all $i \in [1..n]$ 

\[
\mbox{each $s_i \in S_i \setminus S'_i$ is $R$-dominated in $G$ by some $m'_i \in M'_i$}.
\]
\end{itemize}

So, as before, the difference between the $\myra_{\hspace{-1mm} R}$
and $\Ra_{\hspace{-1mm} R}$ lies in the requirement we put on the
$R$-dominating ---this time mixed--- strategy.  In
$\myra_{\hspace{-1mm} R}$ we require that each strategy removed
from $S_i$ is $R$-dominated in $G$ by a mixed strategy in $M_i$,
while in $\Ra_{\hspace{-1mm} R}$ we require that it is $R$-dominated in
$G$ by a mixed strategy in $M'_i$. So in the latter case no strategy
from the support of the $R$-dominating mixed strategy should be
removed at the same time.

To establish equivalence between both reduction relations we need a
counterpart of the notion of a strict partial order.  Below, we
occasionally write each mixed strategy $m'$ over the set of strategies
$S_i$ as the sum $\sum_{t \in S_i} p_t \: t$, where each $p_t = m'(t)$.
Then given two mixed strategies $m_{1}, m_{2}$ and a strategy $t_1$ we
mean by $m_{2}[t_1/m_1]$ the mixed strategy obtained from $m_{2}$ by
substituting the strategy $t_1$ by $m_1$ and by `normalizing' the
resulting sum.

We now say that a mixed dominance relation $R$ is \oldbfe{regular} if
in every game
\begin{itemize}
\item for all $\alpha \in (0,1]$, 
$s \ R \ (1 - \alpha) s + \alpha \: m$ implies
$s \ R \ m$,

\item $t_1 \ R \ m_1$ and $t_2 \ R \ m_2$ implies
$t_1 \ R \ m_{1}[t_2/m_2]$.

\end{itemize}

\begin{lemma}[Equivalence] \label{lem:equm}
For a mixed dominance relation $R$ that is regular
the relations $\myra_{\hspace{-1mm} R}$  and $\Ra_{\hspace{-1mm} R}$ coincide.
\end{lemma}
\Proof
We only need to show that $G \myra_{\hspace{-1mm} R \: } G'$
implies $G \Ra_{\hspace{-1mm} R \: } G'$.

Let $G := (S_1, \LL, S_n,$ $p_1, \LL, p_n)$ and $G' := (S'_1, \LL,
S'_n, p_1, \LL, p_n)$.  Take some $s''_i \in S_i \setminus S'_i$.  Let
$S_i \setminus S'_i := \{t_1, \LL, t_k\}$ with $t_k = s''_i$.  By
definition for all $j \in [1..n]$ some $m_j \in M_i$ exists such that
$t_j \ R \ m_j$ (holds in $G$).
We prove by complete induction that in fact for all $j \in [1..k]$
some $m'_j \in M_i$ exists such that $t_j \ R \ m'_j$ and
$support(m'_{j}) \cap \C{t_1, \LL, t_{j}} = \ES$.

For some $\alpha \in (0,1]$ and a mixed strategy $m'_1$ with $t_1 \not
\in support(m'_1)$ we have
\[
m_1 = (1 - \alpha) t_1 + \alpha \: m'_1.
\]
Since $R$ is regular, $t_1 \ R \ m_1$ implies $t_1 \ R \ m'_1$,
which proves the claim for $k = 1$.

Assume now the claim holds for all $\ell \in [1..j]$.
We have $t_{j+1} \ R \ m_{j+1}$. As in the case of $k = 1$ a mixed strategy
$m''_{j+1}$ exists such that $t_{j+1} \not \in support(m''_{j+1})$ and 
$t_{j+1} \ R \ m''_{j+1}$. Let

\[
m'_{j+1} := m''_{j+1} [t_1/m'_1] \LL [t_j/m'_j]. 
\]
Then for all $\ell \in [1..j]$ we have $support(m''_{j+1} [t_1/m'_1]
\LL [t_{\ell}/m'_{\ell}]) \cap \C{t_1, \LL, t_{\ell}, t_{j+1}} = \ES$,
so $support(m'_{j+1}) \cap \C{t_1, \LL, t_{j+1}} = \ES$, i.e.,
$support(m'_{j+1}) \sse S'_i$. 

Also $t_{j+1} \ R \ m''_{j+1}$ and
$t_{\ell} \ R \ m'_{\ell}$ for all $\ell \in [1..j]$ imply 
by the regularity of $R$ that $t_{j+1} \ R \ m'_{j+1}$.
Hence $s''_i$ (which equals $t_k$)
is $R$-dominated by the mixed strategy $m'_k \in M'_i$.
\HB
\VV

The second condition of the regularity notion appears in Lemma 1
of \cite{Rob03} under the name `transitivity'. In that paper order
independence of conditional dominance is established, a notion
introduced in \cite{SW98}.  Establishing `transitivity' for a
specialized form of conditional dominance (called a robust
demi-replacement) turns out to be a crucial step in the proof
of the order independence.
In our case regularity allows us to focus our representation on the second reduction
relation, $\Ra_{\hspace{-1mm} R}$.

In analogy to the case of dominance relations we say that a mixed dominance relation
$R$ is \oldbfe{hereditary} if for every game $G$, its restriction $G'$,
a strategy $s_i$ of $G'$ and a mixed strategy $m'_i$ of $G'$
\[
\mbox{$s_i$ is $R$-dominated by $m'_i$ in $G$ implies $s_i$ is $R$-dominated by $m'_i$ in $G'$.}
\]

Also, as in the case of the dominance relations, given a mixed
dominance relation $R$ we can specialize the reduction relation
$\Ra_{\hspace{-1mm} R}$ to $\Ra_{\hspace{-1mm} 1, R}$ in which a
single strategy is removed.
The following counterpart of the One-at-a-time Elimination Theorem \ref{thm:one} 
then holds.

\begin{theorem}[One-at-a-time Elimination] \label{thm:one2}
  For a mixed dominance relation $R$ that is hereditary the
  $\Ra_{\hspace{-1mm} R}$ satisfies the one-at-a-time property.
\end{theorem}

\Proof Analogous to the proof of the One-at-a-time Elimination Theorem
\ref{thm:one} and left to the reader.  
\HB 
\VV

As in Section \ref{sec:dominance} for
a mixed dominance relation $R$ that is hereditary and regular we have
three ways of proving that the reduction relation $\Ra_{\hspace{-1mm} R}$
is UN. Here, for a change, we provide a proof for the first approach.

\begin{lemma}[One Step Closedness] \label{thm:onem}
  For a mixed dominance relation $R$ that is hereditary and regular the $\Ra_{\hspace{-1mm} R}$
relation is one step closed.
\end{lemma}
\Proof
Given a game 
$
G := (S_1, \LL, S_n, p_1, \LL, p_n),
$
let
$
G'' := (S''_1, \LL, S''_n,$  $p_1, \LL, p_n)
$
be the game obtained from $G$ by removing all the strategies that are $R$-dominated 
by a mixed strategy in $G$.
Then $G \myra^{\epsilon}_{\hspace{-1mm} R \: } G''$, so
by the Equivalence Lemma \ref{lem:equm} $G \Ra^{\epsilon}_{\hspace{-1mm} R \: } G''$.

Suppose now that $G \Ra_{\hspace{-1mm} R \: } G'$ for some 
$
G' := (S'_1, \LL, S'_n, p_1, \LL, p_n).
$
Then clearly $S''_i \sse S'_i$ for all $i \in [1..n]$.
If $G'$ and $G''$ coincide, then 
$G' \Ra^{\epsilon}_{\hspace{-1mm} R \: } G''$.

Otherwise fix $i \in [1..n]$ and consider a strategy $s_i$ such that $s_i \in
S'_i \setminus S''_i$.
So $s_i$ is eliminated in
the step $G \Ra_{\hspace{-1mm} R \: } G''$. Hence
$s_i$ is $R$-dominated in $G$ by a mixed strategy $m'_i \in M''_i$.
By the hereditarity of $R$ \ 
$s_i$ is $R$-dominated in $G'$ by $m'_i$.
This proves
$G' \Ra_{\hspace{-1mm} R \: } G''$.
\HB
\VV

The reader may note a `detour' in this proof through the $\myra$ reduction, 
justified by the Equivalence Lemma \ref{lem:equm}.
The above lemma brings us to the following conclusion.

\begin{theorem}[Mixed Elimination] \label{thm:strictmixed}
For a mixed dominance relation $R$ that is hereditary and regular
the $\Ra_{\hspace{-1mm} R}$ relation is UN.
\end{theorem}
\Proof 
We noted already in Section \ref{sec:ars} that one step
closedness implies weak confluence. So Newman's Lemma \ref{lem:newman}
applies.  
\HB 
\VV

In other words, when $R$ is a mixed dominance relation that is
hereditary and regular, the process of iterated elimination of
$R$-dominated strategies yields a unique outcome.

We can directly apply the results of this section to
strict dominance by mixed strategies.
Let us recall first the definition.
Consider a game $(S_1, \LL, S_n,$  $p_1, \LL,
p_n)$.  We say that a 
strategy $s_i$ is \oldbfe{strictly dominated} by a mixed strategy
$m'_i$, or equivalently, that 
a mixed strategy $m'_i$ \oldbfe{strictly
  dominates} a strategy $s_i$, if
\[
p_i(s_i, s_{-i}) < p_i(m'_i, s_{-i})
\]
for all $s_{-i} \in S_{-i}$.

This mixed dominance relation entails the reduction relation
$\Ra_{\hspace{-1mm} SM}$ on games obtained by
instantiating the mixed dominance relation $R$ in 
$\Ra_{\hspace{-1mm} R}$ 
by the strict dominance in the above sense.
Clearly, strict dominance by a mixed strategy is hereditary and regular, so
by virtue of the above results  we get the following
counterpart of the Strict Elimination Theorem \ref{thm:strict1}.

\begin{theorem}[Strict Mixed Elimination] \label{thm:strictm}
\mbox{} \vspace{-3mm} 
  \begin{enumerate} \smallromani

\item 
The  $\Ra_{\hspace{-1mm} SM}$ relation is UN.

\item 
The  $\Ra_{\hspace{-1mm} SM}$ relation satisfies the one-at-a-time property.
\HB
\end{enumerate}
\end{theorem}

The first item states that strict dominance by means of
mixed strategies is order independent.

\section{Mixed Strategies: Inherent Dominance}
\label{sec:borm}

The concepts and results of Section \ref{sec:bor} can be naturally
modified to the case of mixed dominance relations.  Consider such a
relation $R$ and a game $(S_1, \LL, S_n,$ $p_1, \LL, p_n)$ and let
$\tilde{S}_{-i}$ be a non-empty subset of $S_{-i}$.  We say that a
strategy $s_i$ is $R$-\oldbfe{dominated given $\tilde{S}_{-i}$} by a
mixed strategy $m'_i$ if $s_i$ is $R$-dominated by $m'_i$ in the
game $(S_i, \tilde{S}_{-i}, p_1, \LL, p_n)$ and say that a strategy
$s_i$ is \oldbfe{inherently $R$-dominated} if for every non-empty
subset $\tilde{S}_{-i}$ of $S_{-i}$ it is $R$-dominated given
$\tilde{S}_{-i}$ by some mixed strategy $m'_i$.

As before, each mixed dominance relation $R$ entails two reduction
relations $\myra_{\hspace{-1mm} inh-R}$ and $\Ra_{\hspace{-1mm}
  inh-R}$ on games  and their `one-at-a-time' versions,
$\myra_{\hspace{-1mm} 1, inh-R}$ and $\Ra_{\hspace{-1mm} 1, R}$.

The \oldbfe{individual independence of irrelevant alternatives} condition 
(\oldbfe{IIIA}) now holds for a mixed dominance relation $R$
if for every game $(S_i, S_{-i},$  $p_1, \LL, p_n)$
\[
  \begin{array}{l}
\mbox{for all $i \in [1..n]$, all non-empty $S'_i \sse S_i$, $s_i \in S'_i$ and $m_i \in M'_i$} \\
\mbox{$s_i \ R \ m'_i$ holds in $(S_i, S_{-i}, p_1, \LL, p_n)$ iff it
 holds in $(S'_i, S_{-i}, p_1, \LL, p_n)$.}
  \end{array}
\]

By analogy we obtain the following results concerning the introduced reduction relations.

\begin{lemma}[Equivalence] \label{lem:equdm}
For a mixed dominance relation $R$ that is regular
the relations $\myra_{\hspace{-1mm} inh-R}$  and $\Ra_{\hspace{-1mm} inh-R}$ coincide.
\end{lemma}
\Proof Analogous to the proof of the Equivalence Lemma \ref{lem:equm} and omitted.
\HB

\begin{lemma}[One Step Closedness] \label{thm:onedm}
  For a mixed dominance relation $R$ that satisfies the IIIA condition
  and is regular the $\Ra_{\hspace{-1mm} inh-R}$ relation is one
  step closed.
\end{lemma}
\Proof Analogous to the proof of the One Step Closedness Lemma \ref{thm:onem},
using the Equivalence Lemma \ref{lem:equdm}, and omitted.
\HB

\begin{theorem}[Inherent Mixed Elimination] \label{thm:strict4}
For a mixed dominance relation $R$ that satisfies the IIIA condition and is regular
the $\Ra_{\hspace{-1mm} \: inh-R \: }$ relation is UN.
\HB
\end{theorem}
\Proof By the One Step Closedness Lemma \ref{thm:onedm}
and Newman's Lemma \ref{lem:newman}.
\HB

\begin{theorem}[One-at-a-time Elimination] \label{thm:oneBd}
  For a mixed dominance relation $R$ that satisfies the IIIA condition
  the $\Ra_{\hspace{-1mm} inh-R}$ relation satisfies the one-at-a-time
  property.
\end{theorem}

\Proof
Analogous to the proof of the One-at-a-time Elimination Theorem \ref{thm:one} and omitted.
\HB
\VV

These results can be directly applied to weak dominance by a mixed
strategy.  Recall that given a game $(S_1, \LL, S_n, p_1, \LL, p_n)$
we say that a strategy $s_i$ is \oldbfe{weakly dominated} by a mixed
strategy $m'_i$, and write $s_i \ {\it WM} \ m'_i$, if
\[
p_i(s_i, s_{-i}) \leq p_i(m'_i, s_{-i})
\]
for all $s_{-i} \in S_{-i}$, with some disequality being strict.

It is straightforward to check that
\emph{WM} satisfies the IIIA condition and is regular. However,
somewhat unexpectedly, we do not get now any new results, since
as shown by \cite{Bor90} the reduction relations $\myra_{\hspace{-1mm}
  inh-WM}$ and $\myra_{\hspace{-1mm} SM}$ 
(and hence $\Ra_{\hspace{-1mm}
  inh-WM}$ and $\Ra_{\hspace{-1mm} SM}$)
coincide.

\section{More on Abstract Reduction Systems}
\label{ars-more}

We shall soon deal with the elimination of payoff equivalent
strategies and to this end we shall need a refinement of 
Newman's Lemma \ref{lem:newman}.  Consider an abstract
reduction system $(A,\myra)$ and assume an equivalence relation
$\sim$ on $A$. We now relativize the previously introduced notions
to $\sim$ and introduce one new concept linking $\myra$ and $\sim$.

\begin{itemize}

\item If every element of $A$ has a unique up to $\sim$ normal form, we say that
$(A,\myra)$ (or simply $\myra$)
satisfies the \oldbfe{$\sim$-unique normal form property}.

\item We say that $\myra$ is \oldbfe{$\sim$-weakly confluent}
if for all $a,b,c \in A$ 

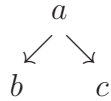
\begin{center}
$a$                                             \\
$\swarrow$ $\searrow$                           \\
$b$\ \ \ \ \ \ \ $c$                            
\end{center}

\NI
implies that for some $d_1, d_2 \in A$

\begin{center}
$b$\ \ \ \ \ \ \ \ \ \ \ \ \ \ $c$                        \\
$\searrow \!*$ \ \ \ \ \ $\!*\swarrow$                    \\
$d_1 \sim d_2$
\end{center}

\item We say that $\myra$ is \oldbfe{$\sim$-bisimilar} if
for all $a,b,c \in A$ 
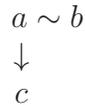
\begin{center}
\ \ \ \ \ $a \sim b$                                             \\
$\downarrow$ \\
$c$ 
\end{center}

\NI
implies that for some $d \in A$

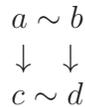
\begin{center}
\ \ \ \ \ $a \sim b$                                             \\
\ \ \ \ \ $\downarrow$ \ \ $\downarrow$ \\
\ \ \ \ \ $c \sim d$ 
\end{center}

\end{itemize}

The following lemma is then a relativized version of 
Newman's Lemma \ref{lem:newman}. 
It is a special case of Lemma 2.7 from
\cite[ page 803]{Hue80}, with a more direct proof.

\begin{lemma}[$\sim$-Newman] \label{lem:newman2}
Consider an abstract reduction system $(A,\myra)$ and an equivalence
relation $\sim$ on $A$ such that

\begin{itemize}
\item no infinite $\myra$ sequences exist,

\item $\myra$ is $\sim$-weakly confluent,

\item $\myra$ is $\sim$-bisimilar.

\end{itemize}
Then $\myra$ satisfies the $\sim$-unique normal form property.
\end{lemma}

\Proof 
We modify the proof of Newman's Lemma \ref{lem:newman}.
We call now an element $a$ \emph{ambiguous} if it
has at least two normal forms that are not equivalent w.r.t.~$\sim$. 
As before  we show that for
every ambiguous $a$ some ambiguous $b$ exists such that $a \myra b$.
This proves absence of ambiguous elements by the first assumption.

So suppose that some element $a$ has two distinct normal forms $n_1$
and $n_2$ such that $n_1 \not\sim n_2$.
Then for some $b, c$ we have $a \myra b \tra n_1$ and $a
\myra c \tra n_2$.  By the $\sim$-weak confluence some $d_1$ and $d_2$ exist such that $b
\tra d_1$, $c \tra d_2$ and $d_1 \sim d_2$. Let $n_3$ be a normal form of $d_1$. Then it
is a normal form of $b$, as well.

By the repeated use of the $\sim$-bisimilarity of $\myra$

\begin{center}
\ \ \ \ \ $d_1 \sim d_2$                                             \\
$\downarrow _{*}$ \\
$n_3$ 
\end{center}

\NI
implies that for some $n_4 \in A$

\begin{center}
\ \ \ \ \ $d_1 \sim d_2$                                             \\
\ \ \ \ \ $\downarrow _{*}$ \ \ \ $\downarrow _{*}$ \\
\ \ \ \ \ $n_3 \sim n_4$ 
\end{center}

Since $n_3$ is a normal form, by the $\sim$-bisimilarity of $\myra$ so is $n_4$.
So $n_4$ is a normal form of $c$.
Moreover $n_3 \not\sim n_1$ or $n_3 \not\sim n_2$, since otherwise $n_1 \sim n_2$ 
would hold. If $n_3 \not\sim n_1$, then $b$ is ambiguous and $a \myra b$.  
And if $n_3 \not\sim n_2$, then also $n_4 \not\sim n_2$
and then $c$ is ambiguous and $a \myra c$.  
\HB
\VV

Also, we have the following relativized version of 
the Unique Normal Form Note \ref{lem:uni}.

\begin{note} [$\sim$-Unique Normal Form] \label{lem:disp:4}
Consider two abstract reduction systems $(A, \myra_{\hspace{-1mm}
  1 \: })$ and $(A, \myra_{\hspace{-1mm} 2 \: })$ and an equivalence
relation $\sim$ on $A$ such that

\begin{itemize}
\item $\myra_{\hspace{-1mm} 1 \: }$ satisfies the $\sim$-unique normal form property,

\item $\myra_{\hspace{-1mm} 1}^{\hspace{-1mm} +} = \myra_{\hspace{-1mm} 2}^{\hspace{-1mm} +}$.

\end{itemize}
Then $\myra_{\hspace{-1mm} 2}$ satisfies the $\sim$-unique normal form property.
\HB
\end{note}

We shall also study the combined effect of two forms of elimination. In what follows we
abbreviate  $\myra_{\hspace{-1mm} 1\: } \cup \myra_{\hspace{-1mm} 2 \: }$ 
to $\myra_{\hspace{-1mm} 1 \vee 2\: }$.
(The use of $\cup$ instead of $\vee$ would clash with the notation used in Section 
\ref{sec:combining}.)
Given two abstract reduction systems $(A, \myra_{\hspace{-1mm} 1 \:
  })$ and $(A, \myra_{\hspace{-1mm} 2 \: })$ we say that
$\myra_{\hspace{-1mm} 1}$ \oldbfe{left commutes with}
$\myra_{\hspace{-1mm} 2}$ if
\[
\myra_{\hspace{-1mm} 1 \: } \circ \myra_{\hspace{-1mm} 2 \: } \sse 
\myra_{\hspace{-1mm} 2 \: } \circ \myra^{\hspace{-1mm} *}_{\hspace{-1mm} 1 \: },
\]
i.e., if for all $a,b,c \in A$ \ 
$a \myra_{\hspace{-1mm} 1 \: } b \myra_{\hspace{-1mm} 2 \: } c$
implies that for some $d \in A$ \ 
$a \myra_{\hspace{-1mm} 2 \: } d \myra^{\hspace{-1mm} *}_{\hspace{-1mm} 1 \: } c$.

\begin{note}[Left Commutativity] \label{not:left} 
If $\myra_{\hspace{-1mm} 1}$ left commutes with $\myra_{\hspace{-1mm} 2}$,
then so does  $\myra^{\hspace{-1mm} +}_{\hspace{-1mm} 1}$.
\HB
\end{note}

Then we shall rely on the following result.

\begin{lemma}[Normal Form] \label{lem:nf}
Consider two abstract reduction systems 

\NI
$(A, \myra_{\hspace{-1mm}
  1 \: })$ and $(A, \myra_{\hspace{-1mm} 2 \: })$ and an equivalence
relation $\sim$ on $A$ such that

\begin{itemize}
\item $(A,\myra_{\hspace{-1mm} 1 \vee 2})$ satisfies the $\sim$-unique normal form property,

\item $\myra_{\hspace{-1mm} 1}$ left commutes with $\myra_{\hspace{-1mm} 2}$.
\end{itemize}
Then for all $a \in A$, if
  
\begin{center}
$a$                                             \\
$*\hspace{-1mm}\swarrow\hspace{-2mm}_{2}$ $\hspace{2mm}_{2}\hspace{-2mm}\searrow \hspace{-1mm}*$                           \\
$b$\ \ \ \ \ \ \ \ $c$                            \\
\end{center}
for some $\myra_{\hspace{-1mm} 2 \: }$-normal forms
$b$ and $c$, then for some $\myra_{\hspace{-1mm} 1 \vee 2 \: }$-normal forms $d_1, d_2 \in A$

\begin{center}
\ $b$ \ \ \ \ \ \ \ \ \  \  $c$                        \\
$\hspace{2mm}_{1}\hspace{-2mm}\searrow \hspace{-1mm}* $ \ \ \ \ \ $* \hspace{-1mm}\swarrow\hspace{-2mm}_{1}$                    \\
$\ \ d_1 \sim d_2$.
\end{center}
\end{lemma}
\Proof
Suppose that $a \myra^{\hspace{-1mm} *}_{\hspace{-1mm} 2} b$ and $a \myra^{\hspace{-1mm} *}_{\hspace{-1mm} 2} c$
where $b$ and $c$ are $\myra_{\hspace{-1mm} 2 \: }$-normal forms.
By the first assumption for some
$\myra_{\hspace{-1mm} 1 \vee 2 \: }$-normal forms $d_1, d_2 \in A$ we have 
$b \myra^{\hspace{-1mm} *}_{\hspace{-1mm} 1 \vee 2} d_1$, $c \myra^{\hspace{-1mm} *}_{\hspace{-1mm} 1 \vee 2} d_2$
and $d_1 \sim d_2$.

If for some $e_1, e_2 \in A$ we have 
$b \myra^{\hspace{-1mm} +}_{\hspace{-1mm} 1} e_1 \myra_{\hspace{-1mm} 2} e_2
 \myra^{\hspace{-1mm} *}_{\hspace{-1mm} 1 \vee 2} d_1$, then by the second assumption 
and the Left Commutativity Note \ref{not:left}
for some $e_3 \in A$ we have $b \myra_{\hspace{-1mm} 2} e_3 \myra^{\hspace{-1mm} *}_{\hspace{-1mm} 1} e_2$,
which contradicts the choice of $b$. So in the path $b \myra^{\hspace{-1mm} *}_{\hspace{-1mm} 1 \vee 2} d_1$
there are no $\myra_{\hspace{-1mm} 2}$ transitions.
By the same argument also in the path $c \myra^{\hspace{-1mm} *}_{\hspace{-1mm} 1 \vee 2} d_2$
there are no $\myra_{\hspace{-1mm} 2}$ transitions.
\HB

\section{Pure Strategies: Payoff Equivalence}
\label{sec:pure-equ}

We now move on to a study of the elimination of payoff equivalent strategies.
This binary relation on the strategies, {\it PE}, entails
the corresponding reduction relation $\Ra_{\hspace{-1mm} {\it PE}}$ on the
games. Let us recall the definition.  Given a game $G := (S_1, \LL, S_n,
p_1, \LL, p_n)$ and its restriction $G' := (S'_1, \LL, S'_n, p_1, \LL,
p_n)$

\begin{itemize}

\item 
$G \Ra_{\hspace{-1mm} {\it PE} \: } G'$ iff $G \neq G'$ and for all $i \in [1..n]$ 

\[
\mbox{each $s_i \in S_i \setminus S'_i$ is payoff equivalent in $G$ to some $s'_i \in S'_i$}.
\]
\end{itemize}

Note that $\Ra_{\hspace{-1mm} {\it PE}}$ is not weakly confluent and
it does not satisfy the unique normal form property. Indeed, given two
payoff equivalent strategies $r$ and $s$, the removal of $r$ and the
removal of $s$ yields two different games. But these games are
obviously equivalent in the sense that a renaming of their strategies
makes them identical.  To study the effect of the removal of the
payoff equivalent strategies we shall therefore consider the following
\oldbfe{equivalence relation} $\sim$ between two games, $G := (S_1,
\LL, S_n, p_1, \LL, p_n)$ and $G' := (S'_1, \LL, S'_n, p'_1, \LL,
p'_n)$:

\[
  \begin{array}{l}
\mbox{$G' \sim G''$ iff for all $i \in [1..n]$ there exists a 1-1 and onto mapping $f_i : S_i \myra S'_i$} \\
\mbox{such that for all $i \in [1..n]$ and $s_i \in S_i$, \ $p_i(s_1, \LL, s_n) = p'_i(f_1(s_1), \LL, f_n(s_n))$.}
  \end{array}
\]

In what follows we shall consider various (also mixed) reduction relations
$\Ra_{\hspace{-1mm} R}$ on games in presence of the $\sim$ equivalence
relation on the games. In each case it will be straightforward to see
that $\Ra_{\hspace{-1mm} R}$ is $\sim$-bisimilar. 
Intuitively, the $\sim$-bisimilarity of $\Ra_{\hspace{-1mm} R}$
simply means that $R$ does not depend
on the strategy names.

Note that if a (mixed) reduction relation
$R$ is hereditary, then to prove that $\Ra_{\hspace{-1mm} R}$ is $\sim$-bisimilar
it is sufficient on the account of the One-at-a-time
Elimination Theorems \ref{thm:one} and \ref{thm:one2}
to check that $\Ra_{\hspace{-1mm} 1,  R}$ is $\sim$-bisimilar.

Instead of saying that a reduction relation
$\Ra_{\hspace{-1mm} R}$ on the set of all restrictions of a game $H$ satisfies the
$\sim$-unique normal form property, we shall simply say that $\Ra_{\hspace{-1mm} R}$
\oldbfe{is $\sim$-UN}.

To reason about the $\Ra_{\hspace{-1mm} {\it PE} \: }$ reduction relation we
shall focus on the relation $\Ra_{\hspace{-1mm} 1, {\it PE}}$ concerned with
the removal of a single strategy payoff equivalent strategy.  The
following simple observation holds.

\begin{lemma}[Weak Confluence] \label{lem:diamond4}
  Consider a game $H$. The $\Ra_{\hspace{-1mm} 1, {\it PE}}$ relation
  on the set of all restrictions of a game $H$ is $\sim$-weakly confluent.
\end{lemma}

\Proof Suppose $G \Ra_{\hspace{-1mm} 1, {\it PE} \: } G'$ and $G
\Ra_{\hspace{-1mm} 1, {\it PE} \: } G''$.  Let $r$ and $s$ be the
strategies eliminated in the first, respectively second, transition.
If $r$ and $s$ are payoff equivalent in $G$, then $G' \sim G''$.
Otherwise, by the hereditarity of {\it PE}, $G' \Ra_{\hspace{-1mm} 1,
  {\it PE} \: } G' \cap G''$ and $G'' \Ra_{\hspace{-1mm} 1, {\it PE}
  \: } G' \cap G''$.  
\HB 
\VV

This brings us to the following result that we shall need in the sequel.

\begin{theorem}[Payoff Equivalence Elimination] \label{thm:elim}
\mbox{} \vspace{-3mm} 
  \begin{enumerate} \smallromani

  \item The $\Ra_{\hspace{-1mm} 1, {\it PE}}$ relation
  is $\sim$-UN.

  \item The $\Ra_{\hspace{-1mm} {\it PE}}$ relation is $\sim$-UN.
  \end{enumerate}
\end{theorem}
\Proof

\NI 
$(i)$ We just proved that $\Ra_{\hspace{-1mm} 1, PE}$ is
$\sim$-weakly confluent.  Also, this reduction relation is clearly
$\sim$-bisimilar.  So the conclusion follows by the $\sim$-Newman's
Lemma \ref{lem:newman2}.

\NI
$(ii)$
First note that {\it PE} is hereditary, so
by the One-at-a-time Elimination Theorem \ref{thm:one}
$\Ra_{\hspace{-1mm} PE}$ satisfies the one-at-a-time property,
that is, 
\[
\Ra^{\hspace{-1mm} +}_{\hspace{-1mm} 1, PE} = \Ra^{\hspace{-1mm} +}_{\hspace{-1mm} PE}.
\]
It suffices now to apply the $\sim$-Unique Normal Form Note
\ref{lem:disp:4}.
\HB
\VV

Informally, the process of iterated elimination of payoff equivalent
strategies yields a unique outcome up to the introduced equivalence
relation $\sim$ on the games.  This outcome can also be achieved in
one step, by replacing each maximal set of at least two mutually
payoff equivalent strategies by one representative. The resulting game
is called in \cite{Mye91} a \oldbfe{purely reduced} game.  Of course,
the above result is completely expected.  Still, we find that a
concise formal justification of it is in order.

\section{Mixed Strategies: Randomized Redundance}
\label{sec:randomized}

The notion of payoff equivalent strategies generalizes in the obvious
way to the mixed strategies. We denote by \emph{PEM} the corresponding
mixed dominance relation. So for a strategy $s_i$ and a mixed
strategy $m'_i$ of player $i$ \ $s_i \: \emph{PEM} \ m'_i$ if
\[
p_j(s_i, s_{-i}) = p_j(m'_i, s_{-i})
\]
for all $j \in [1..n]$ and all $s_{-i} \in S_{-i}$.

As explained in Section \ref{sec:mixeddom} \emph{PEM} entails the
reduction relation $\Ra_{\hspace{-1mm} {\it PEM}}$ 
on games.
Recall that a strategy $s_i$ of player $i$ is called
\oldbfe{randomized redundant to} a mixed strategy $m_i$ if it is payoff equivalent to 
$m_i$ and $s_i \not \in support(m_i)$.
Note that for a game $(S_1, \LL, S_n, p_1, \LL, p_n)$ and its restriction
$G' := (S'_1, \LL, S'_n,$  $p_1, \LL, p_n)$ we have

\begin{itemize}
\item 
$G \Ra_{\hspace{-1mm} {\it PEM} \: } G'$ when $G \neq G'$ and for all $i \in [1..n]$ 

\[
\mbox{each $s'_i \in S_i \setminus S'_i$ is randomized redundant in $G$ to some $m'_i \in M'_i$}.
\]
\end{itemize}

As in the case of payoff equivalence it is
more convenient to focus on the removal of a single strategy, so on the reduction 
relation $\Ra_{\hspace{-1mm} 1, {\it PEM}}$.
The following counterpart of the Weak Confluence Lemma
\ref{lem:diamond4} holds.

\begin{lemma}[Weak Confluence] \label{lem:diamond5}
Consider a game $H$. The  $\Ra_{\hspace{-1mm} 1, {\it PEM}}$ relation on the set of all restrictions
of a game $H$ is $\sim$-weakly confluent.
\end{lemma}
\Proof
Suppose $G \Ra_{\hspace{-1mm} 1, {\it PEM} \: } G'$ and $G
\Ra_{\hspace{-1mm} 1, {\it PEM} \: } G''$.  Let $r$ and $t$ be the
strategies eliminated in the first, respectively second, transition.
If $r$ and $t$ are payoff equivalent, then, as in the proof of the Weak Confluence Lemma
\ref{lem:diamond4}, $G' \sim G''$.

Otherwise for some $\alpha \in [0,1)$ and $\beta \in [0,1)$
$r$ is payoff equivalent to a mixed strategy $\alpha \: t + (1 - \alpha) m_1$ with $r,t \not \in support(m_1)$
and $t$ is payoff equivalent to a mixed strategy $\beta \: r + (1 - \beta) m_2$ with $r,t \not \in support(m_2)$.
So $r$ is payoff equivalent to $\alpha \: \beta \: r + \alpha (1 - \beta) m_2 + (1 - \alpha) m_1$,
and hence to 
\[
m' := (\alpha (1 - \beta) m_2 + (1 - \alpha) m_1) / (1- \alpha \: \beta).
\]
Since $t \not \in support(m')$, $m'$ is a mixed strategy in $G''$. So
by the hereditarity of \emph{PEM} \ $r$ is payoff equivalent to $m'$ in $G''$.
Further, since $r,t \not \in support(m')$, $m'$ is a mixed strategy in $G' \cap G''$.
So we showed that $G'' \Ra_{\hspace{-1mm} 1, {\it PEM} \: } G' \cap G''$.  
By symmetry $G' \Ra_{\hspace{-1mm} 1, {\it PEM} \: } G' \cap G''$.  
\HB
\VV

As in the case of the $\Ra_{\hspace{-1mm} {\it PE}}$ relation we can now conclude.

\begin{theorem}[Redundance Elimination] \label{thm:elim2}
\mbox{} \vspace{-3mm} 
  \begin{enumerate} \smallromani
  \item The $\Ra_{\hspace{-1mm} 1, {\it PEM}}$ relation is $\sim$-UN.

  \item 
The $\Ra_{\hspace{-1mm} {\it PEM}}$ relation is $\sim$-UN.
\HB
 \end{enumerate}
\end{theorem}

So the process of iterated elimination of randomized redundant
strategies yields a unique up to $\sim$ outcome.
The result is called in \cite{Mye91} a \oldbfe{fully reduced} game.

\section{Combining Two Dominance Relations}
\label{sec:combining}

Given two dominance relation $R, Q$ we now consider the combined
dominance relation $R \cup Q$.  Such a combination is meaningful to
study when $Q$ is such that the $\Ra_{\hspace{-1mm} Q}$ reduction
relation is $\sim$-UN.  An example is the payoff equivalence \emph{PE}
relation discussed in Section \ref{sec:pure-equ}.
  
Given two dominance relations $R$ and $Q$ we would like now to
identify conditions that allow us to conclude that the $\Ra_{\hspace{-1mm}
    R \cup Q}$ reduction relation is $\sim$-UN.
To this end we introduce the following
  concept.  We say that $R$ is \oldbfe{closed
    under} $Q$ if in all games $G$ for all strategies $r,s,t$
\begin{itemize}
\item  $r \ R \ s$ and $s \ Q \ t$ implies $r \ R \ t$,

\item  $r \ Q \ s$ and $s \ R \ t$ implies $r \ R \ t$,

\end{itemize}
i.e., if in all games $R \circ Q \sse R$ and $Q \circ R \sse R$.

Here is a result that we shall use in the sequel.

\begin{theorem}[Combination] \label{thm:combination}
Consider two dominance relations $R$ and $Q$ such that 

\begin{itemize}
\item $\Ra_{\hspace{-1mm} R}$ and $\Ra_{\hspace{-1mm} Q}$ are  $\sim$-bisimilar,

\item $R$ is a strict partial order,

\item $R$ is closed under $Q$,

\item $\Ra_{\hspace{-1mm} \: 1, Q}$ is $\sim$-UN,

\item $R \cup Q$ is hereditary.
\end{itemize}
Then the $\Ra_{\hspace{-1mm} \: R \cup Q}$ relation is $\sim$-UN.
\end{theorem}

Notice that we do not insist here that $R$ is hereditary. In fact, in
one of the uses of the above result the dominance relation $R$ will
not be hereditary.
\II

\Proof 
Since $R \cup Q$ is hereditary, by the
One-at-a-time Elimination Theorem \ref{thm:one} and the
$\sim$-Unique Normal Form Note \ref{lem:disp:4} it suffices
to prove that $\Ra_{\hspace{-1mm} \: 1, R \cup Q}$ is $\sim$-UN.
But by assumption both $\Ra_{\hspace{-1mm} \: 1, R}$ and
$\Ra_{\hspace{-1mm} \: 1, Q}$ are $\sim$-bisimilar, so
$\Ra_{\hspace{-1mm} \: 1, R \cup Q}$ is $\sim$-bisimilar, as well.
So on the account of the $\sim$-Newman's Lemma \ref{lem:newman2} 
the fact that $\Ra_{\hspace{-1mm} \: 1, R \cup Q}$ is $\sim$-UN is
established once we show that $\Ra_{\hspace{-1mm} \: 1, R \cup Q}$ is
$\sim$-weakly confluent.

So suppose that $G \Ra_{\hspace{-1mm} \: 1, R \cup Q} G'$
and $G \Ra_{\hspace{-1mm} \: 1, R \cup Q} G''$.
Let $r$ and $s$ be the
strategies eliminated in the first, respectively second, transition.
By the fourth assumption $\Ra_{\hspace{-1mm} \: 1, Q}$ is
is $\sim$-weakly confluent, so
we only need to consider a situation when
$G \Ra_{\hspace{-1mm} 1,R \: } G'$. 

We can assume that $G' \neq G''$.
Then $r$ is in $G''$ and $s$ is in $G'$.
By definition $r \: R \ t$ holds in $G$ for some strategy $t$ of $G'$
and $s \: R \cup Q \ u$ holds in $G$ for some strategy $u$ of $G''$.
To show that $G'' \Ra_{\hspace{-1mm} \: 1, R \cup Q} G' \cap G''$
we consider two cases.
\II

\NI
\emph{Case 1.} $t$ is in $G''$, i.e., $s \neq t$.

Then, by the hereditarity of $R \cup Q$, $r \ R \cup Q \ t$ holds in $G''$.
\II

\NI
\emph{Case 2.} $t$ is not in $G''$, i.e., $s = t$.

Then $r \: R \ s$ holds in $G$.  If $s \: R \ u$ holds in $G$, then, by the transitivity of $R$
also $r \: R \ u$ holds in $G$.  

If $s \: Q \ u$ holds in $G$, then by the fact that $R$ is closed under $Q$ \ 
$r \: R \ u$ holds in $G$, as well.  Further, $r
\neq u$ by the irreflexivity of $R$, so $u$ is in $G'$.  Hence, by
Case 1, $r \: R \cup Q \ u$ holds in $G''$.  \II

This proves that $G'' \Ra_{\hspace{-1mm} \: 1, R \cup Q} G' \cap G''$.
To show that $G' \Ra_{\hspace{-1mm} \: 1, R \cup Q} G' \cap G''$
we again consider two cases.
\II

\NI
\emph{Case 1.} $u$ is in $G'$, i.e., $u \neq r$.

Then, by the hereditarity of $R \cup Q$, $s \ R \cup Q \ u$ holds in $G'$.
Also $u$ is in $G''$.
\II

\NI
\emph{Case 2.} $u$ is not in $G'$, i.e., $u = r$.

Then $s \: R \cup Q \ r$ holds in $G$.  If $s \: R \ r$ holds in $G$, then, by the
transitivity of $R$, $s \: R \ t$ holds in $G$.  

If $s \: Q \ r$ holds in $G$, then by the fact that $R$ is closed under $Q$ \ 
$s \: R \ t$ holds in $G$, as well.  But $s$ and
$t$ are strategies of $G'$, so by the hereditarity of $R$ \ $s \: R \ 
t$ holds in $G'$.  This shows $G' \myra_{\hspace{-1mm} R \: } G' \cap
G''$.

By the Equivalence Lemma \ref{lem:equ1} the relations
$\myra_{\hspace{-1mm} R}$ and $\Ra_{\hspace{-1mm} R}$ coincide, so
some strategy $t'$ of $G'\cap G''$ exists such that $s \: R \ t'$, and a fortiori $s \:
R \cup Q \ t'$, holds in $G'$.
\II

This proves that $G' \Ra_{\hspace{-1mm} \: 1, R \cup Q} G' \cap G''$.
\HB
\VV

This result is a generalization of the
Elimination Theorem \ref{thm:hereditary}. Indeed, it suffices to
use instead of $\sim$ the identity relation on games, and use as
$Q$ the identity dominance relation (according to which a strategy is only 
dominated by itself). Then the assumptions of the above theorem reduce to those
of the Elimination Theorem \ref{thm:hereditary}. 

As a simple application of this result consider the
combination of the strict dominance and the payoff equivalence.
The strict dominance relation is
hereditary and so is {\it PE}, and a union of two
hereditary dominance relations is hereditary.  Further, strict
dominance is a strict partial order and is easily seen to be closed
under the payoff equivalence.  So the
following direct consequence of the Payoff Equivalence Elimination
Theorem \ref{thm:elim}$(i)$ and of the above result holds.

\begin{theorem}[Combined Strict Elimination] \label{thm:combined}
The  $\Ra_{\hspace{-1mm} S \cup {\it PE} \: }$ relation is $\sim$-UN.
\HB
\end{theorem}

In other words, the combined iterated elimination of strategies in
which at each step we remove some strictly dominated strategies
and some payoff equivalent strategies yields a unique up to the
equivalence relation $\sim$ outcome.

\section{Combining Nice Weak Dominance with
\mbox{\qquad \qquad} Payoff Equivalence}
\label{sec:nweak}

In this section we show another application of the Combination 
Theorem \ref{thm:combination} concerned with a modification of the 
weak dominance. We denote by $\Ra_{\hspace{-1mm} W}$  
the reduction relation on games corresponding to weak dominance.
As mentioned earlier, $\Ra_{\hspace{-1mm} W}$
does not satisfy the unique normal form property.
An example relevant for us will be provided in a moment.

We studied already one modification of weak dominance in Section
\ref{sec:bor} by considering inherent weak dominance, a notion
due to \cite{Bor90}. 
Another approach was pursued in
\cite{MS97} (see also \cite{MS00}) who
studied the notion of nice weak dominance,
introduced in Subsection \ref{subsec:prelim} and denoted by \emph{NW}.
However, the $\Ra_{\hspace{-1mm} {\it NW}}$ reduction relation, 
just as $\Ra_{\hspace{-1mm} W}$, does not satisfy the
unique normal form property.  To see this consider the following game:

\begin{center}
\begin{game}{2}{2}
      & $L$    & $R$\\
$T$   &$2,1$   &$2,1$\\
$B$   &$2,1$   &$1,0$
\end{game}
\end{center}


Clearly, all pairs of strategies are compatible, so weak dominance
and nice weak dominance coincide here.
This game can be reduced by means of the $\Ra_{\hspace{-1mm} {\it NW}}$ relation both to
\begin{center}
\begin{game}{1}{2}
      & $L$    & $R$\\
$T$   &$2,1$   &$2,1$
\end{game}
\end{center}
and to


\begin{center}
\begin{game}{2}{1}
      & $L$  \\
$T$   &$2,1$ \\
$B$   &$2,1$ 
\end{game}
\end{center}


In each case we reached a $\Ra_{\hspace{-1mm} {\it NW}}$-normal form.
So the $\Ra_{\hspace{-1mm} {\it NW}}$ relation (and consequently the
$\Ra_{\hspace{-1mm}  \it{W}}$ relation) is not weakly confluent  
and does not satisfy the unique normal form property.
Note also that the strategy \texttt{L} (nicely) weakly dominates
\texttt{R} in the original game but not in the first first restriction. 
This shows that neither weak dominance nor nice weak dominance is hereditary.

A solution consists of combining nice weak dominance with the
payoff equivalence and seeking conditions under which nice weak
dominance and weak dominance coincide.  This is the approach taken in
\cite{MS97} who proved that the
$\Ra_{\hspace{-1mm} {\it NW}}$-normal forms of a game are the same up
to the removal of the payoff equivalent strategies and a renaming of
strategies.\footnote{Also an addition of payoff equivalent strategies
  is allowed.  Our proof shows this is not needed.}  They also
observed that for the games $(S_1, \LL, S_n, p_1, \LL, p_n)$ that
satisfy the already mentioned in the Introduction \oldbfe{transference of
  decisionmaker indifference} (TDI) condition:

\begin{equation}
  \begin{array}{l}
\mbox{for all $i, j \in [1..n]$, $s'_i, s''_i \in S_i$ and $s_{-i} \in S_{-i}$} \\
\mbox{$p_{i}(s'_i, s_{-i}) = p_{i}(s''_i, s_{-i})$ implies $p_{j}(s'_i, s_{-i}) = p_{j}(s''_i, s_{-i})$,}
  \end{array}
\label{eq:tdi}
\end{equation}
nice weak dominance and weak dominance coincide on all restrictions.  To
see the latter note that the compatibility is hereditary and the TDI
condition simply amounts to a statement that all pairs of strategies
$s'_i$ and $s''_i$ are compatible.  So for the games that satisfy the
TDI condition the $\Ra_{\hspace{-1mm} {\it W}}$-normal forms of a game
are the same up to the removal of the payoff equivalent strategies and
a renaming of strategies.

\cite{MS97} also provided a number of natural examples of games that
satisfy this condition. We now present conceptually simpler proofs of
their results by following the methodology used throughout the paper.
In Section \ref{sec:mnweak} we shall deal with the case of the nice weak
dominance by mixed strategies.

The following lemma summarizes the crucial properties of nice weak dominance.
They are `crucial' in the sense that they allow us to directly apply
the already discussed Combination 
Theorem \ref{thm:combination} to nice weak dominance and payoff equivalence.


\begin{lemma}[Nice Weak Dominance] \label{lem:nice}
\mbox{} \vspace{-3mm} 
  \begin{enumerate} \smallromani

\item 
${\it NW}$ is a strict partial order.

\item ${\it NW}$ is closed under {\it PE}.

\item ${\it NW} \cup {\it PE}$ is hereditary.
  \end{enumerate}
\end{lemma}

\Proof
$(i)$  First, note that the relation ${\it NW}$ is clearly
irreflexive. To prove transitivity consider a game $(S_1, \LL, S_n,
p_1, \LL, p_n)$ and suppose that $s''_i \ {\it NW} \ s'_i$ and $s'_i \ {\it NW} \ s^{*}_i$.

Then clearly $s''_i$ is weakly dominated by $s^{*}_i$.
To prove that $s''_i$ and $s^{*}_i$ are compatible
suppose that for some $s_{-i} \in S_{-i}$
\[
p_{i}(s''_i, s_{-i}) = p_{i}(s^{*}_i, s_{-i}).
\]
Then by the weak dominance
\[
p_{i}(s''_i, s_{-i}) = p_{i}(s'_i, s_{-i}) = p_{i}(s^{*}_i, s_{-i}).
\]
Hence by the compatibility of $s''_i$ and $s'_i$ and the compatibility of 
$s'_i$ and $s^{*}_i$ for all $j \in [1..n]$
\[
p_{j}(s''_i, s_{-i}) = p_{j}(s'_i, s_{-i}) = p_{j}(s^{*}_i, s_{-i}).
\]

\NI
$(ii)$
The proofs of the relevant two properties of ${\it NW}$ are
analogous to the proof of $(i)$ and are omitted. 
\III

\NI
$(iii)$
Let $G' := (S'_1, \LL, S'_n, p_1, \LL, p_n)$ be a restriction of $G := (S_1, \LL, S_n, p_1, \LL, p_n)$.
Suppose $s'_i, s''_i \in S'_i$ are such that $s'_i \ {\it NW} \cup {\it PE} \ s''_i$ in $G$.
Then $s''_i$ and $s'_i$ are
compatible in $G$ and hence in $G'$.
Moreover
\[
p_i(s''_i, s^{*}_{-i}) \geq p_i(s'_i, s^{*}_{-i})
\]
for all $s^{*}_{-i} \in S'_{-i}$.
If for some $s^{*}_{-i} \in S'_{-i}$
\[
p_i(s''_i, s^{*}_{-i}) > p_i(s'_i, s^{*}_{-i}),
\]
then $s''_i$ weakly dominates $s'_i$ in $G'$
and consequently $s''_i$ nicely weakly dominates $s'_i$ in $G'$.
Otherwise 
\[
p_i(s''_i, s^{*}_{-i}) = p_i(s'_i, s^{*}_{-i})
\]
for all $s^{*}_{-i} \in S'_{-i}$, so, by the compatibility of
$s''_i$ and $s'_i$ in $G'$,
$s''_i$ and $s'_i$ are payoff equivalent in $G'$.

So we showed that $s''_i \ {\it NW} \cup {\it PE} \ s'_i$ in $G'$.
\HB
\VV

Nice weak dominance clearly satisfies the IIIA condition
of Section \ref{sec:bor} and by item $(i)$ above it is a strict partial order.
So using $R := \emph{NW}$ in the
Inherent Elimination Theorem \ref{thm:strict3} 
and the One-at-a-time Elimination Theorem \ref{thm:oneB}
we get the following result.

\begin{theorem}[Inherent Nice Weak Elimination] \label{thm:nweak-inh}
\mbox{} \vspace{-3mm} 
  \begin{enumerate} \smallromani

\item 
The $\Ra_{\hspace{-1mm} inh-NW}$ relation is UN.

\item 
The  $\Ra_{\hspace{-1mm} inh-NW}$ relation satisfies the one-at-a-time property.
\HB
\end{enumerate}
\end{theorem}

Further, the above lemma in conjunction with the Payoff Equivalence Elimination
Theorem \ref{thm:elim}$(i)$ means that for $R := \emph{NW}$ and $Q :=\emph{PE}$
all assumptions of the Combination Theorem \ref{thm:combination} 
are satisfied. So we get the following conclusion.

\begin{theorem}[Nice Weak Elimination] \label{thm:nweak1}
The  $\Ra_{\hspace{-1mm} \it{NW}  \cup \it{PE}}$ relation is $\sim$-UN.

\HB
\end{theorem}

Also, for games that satisfy the TDI condition
(\ref{eq:tdi}) the $\Ra_{\hspace{-1mm} \it{NW} \cup \it{PE}}$ and
$\Ra_{\hspace{-1mm} \it{W} \cup \it{PE}}$ relations coincide on all
restrictions, so the following conclusion follows.

\begin{corollary}[Weak Elimination] \label{cor:weak1}
Consider a game $H$ that satisfies the TDI condition (\ref{eq:tdi}). 
Then the  $\Ra_{\hspace{-1mm}  \it{W} \cup \it{PE}}$ relation is $\sim$-UN.
\HB
\end{corollary}

To establish another form of order independence involving nice weak dominance
we shall rely on the following observation that refers to the crucial
concept of left commutativity.

\begin{note}[Left Commutativity] \label{not:left1} 
$\Ra_{\hspace{-1mm} {\it PE} \: }$ left commutes with $\Ra_{\hspace{-1mm} {\it NW}}$.
\end{note}
\Proof
By the One-at-a-time
Elimination Theorem \ref{thm:one} the reduction relation $\Ra_{\hspace{-1mm} PE}$
satisfies the one-at-a-time property, i.e.,  
\[
\Ra^{\hspace{-1mm} +}_{\hspace{-1mm} 1,PE} = \Ra^{\hspace{-1mm} +}_{\hspace{-1mm} PE}. 
\]
So by the Left Commutativity Note \ref{not:left} it suffices to show that
$\Ra_{\hspace{-1mm} 1, {\it PE} \: }$ left commutes with $\Ra_{\hspace{-1mm} {\it NW}}$.
Suppose $G \Ra_{\hspace{-1mm} 1, {\it PE} \: } G' \Ra_{\hspace{-1mm}
  {\it NW}} G''$.  
In the proof below we repeatedly use the fact that if a strategy $r_j$ is nicely
weakly dominated in $G'$ by a strategy $t_j$, then so it is in $G$.

Let $s_i$ be the strategy deleted in the first
transition.  If all strategies that are payoff equivalent to $s_i$ are
removed in the second transition, then by the Nice Weak Dominance
Lemma \ref{lem:nice}\emph{(ii)} ${\it NW}$ is closed under {\it PE} which
implies $G \Ra_{\hspace{-1mm} {\it NW}} G''$. Consequently
$G \Ra_{\hspace{-1mm} {\it NW} \: } G'' \Ra^{\hspace{-1mm}
  \epsilon}_{\hspace{-1mm} 1, {\it PE} \: } G''$.  

Otherwise, by the fact that payoff equivalence is hereditary, we have 

\NI
$G
\Ra_{\hspace{-1mm} {\it NW} \: } G_1 \Ra_{\hspace{-1mm} 1, \it{PE}}
G''$, where $G_1$ is obtained from $G''$ by adding $s_i$ to the set of
strategies of player $i$.  
\HB 
\VV

As an aside, note that the same proof shows that
$\Ra_{\hspace{-1mm} {\it PE} \: }$ left commutes with $\Ra_{\hspace{-1mm} {\it W}}$
and with $\Ra_{\hspace{-1mm} {\it S}}$.
The relevant property is that both ${\it W}$ and ${\it S}$ are
closed under {\it PE}.

We reached now the already mentioned result of \cite{MS97}.

\begin{theorem}[Structured Nice Weak Elimination] \label{thm:weak3}
  Suppose that 

\NI
$G \Ra^{\hspace{-1mm} *}_{\hspace{-1mm} {\it NW} \: } G'$
  and $G \Ra^{\hspace{-1mm} *}_{\hspace{-1mm} {\it NW} \: } G''$, where
  both $G'$ and $G''$
are closed under the $\Ra_{\hspace{-1mm} {\it NW}}$ reduction
(i.e., are $\Ra_{\hspace{-1mm} {\it NW}}$-normal forms).

Then for some $\sim$-equivalent games $H'$ and $H''$
closed under the $\Ra_{\hspace{-1mm} {\it NW} \cup {\it PE}}$ reduction
we have $G' \Ra^{\hspace{-1mm} *}_{\hspace{-1mm} {\it PE} \: } H'$ and 
$G'' \Ra^{\hspace{-1mm} *}_{\hspace{-1mm} {\it PE} \: } H''$.
\end{theorem}

\Proof
Since \emph{PE} is hereditary, each step $H_1 \Ra_{\hspace{-1mm} {\it
    NW} \cup {\it PE} \: } H_2$ can be rewritten as $H_1
\Ra_{\hspace{-1mm} {\it NW} \: } H_3 \Ra_{\hspace{-1mm} {\it PE} \: }
H_2$ for some game $H_3$.
So by the Nice Weak Elimination Theorem \ref{thm:nweak1}
the $\myra_{\hspace{-1mm} 1 \vee 2}$ relation,
where $\myra_{\hspace{-1mm} 1} := \Ra_{\hspace{-1mm} {\it NW}}$ and
$\myra_{\hspace{-1mm} 2} := \Ra_{\hspace{-1mm} {\it PE}}$,
is $\sim$-UN.

It suffices now to use the Left Commutativity Note \ref{not:left1} and
the Normal Form Lemma \ref{lem:nf}. 
\HB
\VV

As explained at the end of Section \ref{sec:pure-equ}
the reductions from $G'$ to $H'$ and from $G''$ to $H''$
can be achieved in just one step.

\begin{corollary}[Structured Weak Elimination] \label{cor:weak3}
Consider a game $G$ that satisfies the TDI condition (\ref{eq:tdi}). 
Suppose that $G \Ra^{\hspace{-1mm} *}_{\hspace{-1mm} W \: } G'$ and 
$G \Ra^{\hspace{-1mm} *}_{\hspace{-1mm} W \: } G''$, where both $G'$ 
and $G''$ are closed under the $\Ra_{\hspace{-1mm} W}$ reduction.

Then for some $\sim$-equivalent games $H'$ and $H''$
we have $G' \Ra^{\hspace{-1mm} *}_{\hspace{-1mm} {\it PE} \: } H'$ and 
$G'' \Ra^{\hspace{-1mm} *}_{\hspace{-1mm} {\it PE} \: } H''$.
\HB
\end{corollary}

Recently, \cite{Ost04} provided an alternative proof of this corollary.

In the Weak Elimination Corollary \ref{cor:weak1} we can weaken the
assumption that the initial game $H$ satisfies the TDI condition.
Indeed, it suffices to ensure that each time an $\Ra_{\hspace{-1mm} {\it W}}$ 
reduction can take place, it is in fact an
  $\Ra_{\hspace{-1mm} {\it NW}}$ reduction. This is guaranteed if the
    following condition TDI$^+$ is satisfied, given an initial game
    $H$:

\[
  \begin{array}{l}
\mbox{for all restrictions $G := (S_1, \LL, S_n, p_1, \LL, p_n)$ of $H$,} \\
\mbox{for all $i \in [1..n]$ and $r_i, t_i \in S_i$} \\
\mbox{if $t_i$ weakly dominates $r_i$ in $G$, then $r_i$ and $t_i$ are compatible in $G$.}
  \end{array}
\]

An alternative, suggested by \cite{MS97} in the context of nice weak
dominance by mixed strategies, is to use the following condition
TDI$^{++}$, where, given a game $(S_1, \LL, S_n, p_1, \LL, p_n)$,
a strategy $s'_i$ \oldbfe{very weakly dominates} a strategy $s''_i$ if
\[
p_i(s'_i, s_{-i}) \geq p_i(s''_i, s_{-i}) 
\]
for all $s_{-i} \in S_{-i}$:
\[
  \begin{array}{l}
\mbox{for all restrictions $G := (S_1, \LL, S_n, p_1, \LL, p_n)$ of $H$,} \\
\mbox{for all $i \in [1..n]$ and $r_i, t_i \in S_i$ if $t_i$ very weakly dominates $r_i$ in $G$, then} \\
\mbox{either $t_i$ weakly dominates $r_i$ in $G$ or $r_i$ and $t_i$ are payoff equivalent in $G$.}
  \end{array}
\]

Indeed, it suffices to show that under the TDI$^{++}$ condition all
assumptions of the Combination Theorem \ref{thm:combination} are
satisfied by the weak dominance relation $W$. First, note that $W$ is
a strict partial order and is clearly closed under the payoff
equivalence.

Denote now the very weak dominance relation by $VW$. Note that
\begin{itemize}
\item $W \sse VW$ (i.e., weak dominance implies very weak dominance),

\item $VW$ is hereditary.

\end{itemize}

Additionally, by the TDI$^{++}$ assumption,

\begin{itemize}
\item $VW \sse W \cup {\it PE}$

\end{itemize}
holds in all restrictions of the initial game $H$.

This implies under the TDI$^{++}$ assumption that $W \cup {\it PE}$ is
hereditary since ${\it PE}$ is hereditary.
By the Combination Theorem \ref{thm:combination} we
conclude then that the $\Ra_{\hspace{-1mm} \it{W} \cup \it{PE}}$
reduction relation is $\sim$-UN.

The same considerations apply to the Structured
Weak Elimination Corollary \ref{cor:weak3}.
However, to be able to use the TDI$^{++}$ condition
we need in addition to prove that $\Ra_{\hspace{-1mm} \it{PE}}$ left commutes with
$\Ra_{\hspace{-1mm} \it{W}}$. The proof is the same as that of the 
Left Commutativity Note \ref{not:left1}.

\section{Combining Two Mixed Dominance Relations}
\label{sec:combiningm}

We now return to the mixed dominance relations
and study a combination $R \cup Q$ of two
such relations $R$ and $Q$. In the applications $Q$ will
be the randomized redundance relation {\it PEM}
studied in Section \ref{sec:randomized}.

We say that a combined mixed dominance relation $R$
is \oldbfe{closed under} $Q$
if in all games $G$ for all strategies $r,s$ and all mixed strategies $m_1, m_2$
\begin{itemize}
\item  $r \ R \ m_1$ and $s \ Q \ m_2$ implies $r \ R \ m_1 [s/m_2]$,

\item  $r \ Q \ m_1$ and $s \ R \ m_2$ implies $r \ R \ m_1 [s/m_2]$.

\end{itemize}

The following counterpart of the Combination 
Theorem \ref{thm:combination} holds. 

\begin{theorem}[Combination] \label{thm:mcombination}
Consider two mixed dominance relations $R$ and $Q$ such that 

\begin{itemize}

\item $\Ra_{\hspace{-1mm} R}$ and $\Ra_{\hspace{-1mm} Q}$ are  $\sim$-bisimilar,

\item $R$ is regular,

\item $R$ is closed under the randomized redundance,

\item $\Ra_{\hspace{-1mm} \: 1, Q}$ is $\sim$-UN,

\item $R \cup Q$ is hereditary.
\end{itemize}
Then the $\Ra_{\hspace{-1mm} \: R \cup Q}$ relation is $\sim$-UN.
\end{theorem}
\Proof We proceed as in the proof of the Combination 
Theorem \ref{thm:combination}.

Since $R \cup Q$ is hereditary, by the
One-at-a-time Elimination Theorem \ref{thm:one} and the
$\sim$-Unique Normal Form Note \ref{lem:disp:4} it suffices
to prove that $\Ra_{\hspace{-1mm} \: 1, R \cup Q}$ satisfies
the $\sim$-unique normal form.  In turn, 
by the $\sim$-Newman's Lemma \ref{lem:newman2}
this is established once we show
that $\Ra_{\hspace{-1mm} \: 1, R \cup Q}$ is $\sim$-weakly
confluent.
Indeed, as before $\Ra_{\hspace{-1mm} \: 1, R \cup Q}$ is $\sim$-bisimilar.

So suppose that $G \Ra_{\hspace{-1mm} \: 1, R \cup Q} G'$
and $G \Ra_{\hspace{-1mm} \: 1, R \cup Q} G''$.
Let $r$ and $s$ be the
strategies eliminated in the first, respectively second, transition.
By the fourth assumption $\Ra_{\hspace{-1mm} \: 1, Q}$ is $\sim$-weakly confluent,
so we only need to consider a situation when
$G \Ra_{\hspace{-1mm} 1,R \: } G'$. 

We can assume that $G' \neq G''$.  Then $r$ is in $G''$ and $s$ is in
$G'$.  By definition $r \: R \ m_1$ holds in $G$ for some mixed
strategy $m_1$ of $G'$ and $s \: R \cup Q \ m_2$ holds in $G$
for some mixed strategy $m_2$ of $G''$.  To show that $G''
\Ra_{\hspace{-1mm} \: 1, R \cup Q} G' \cap G''$ we consider two
cases.  
\II

\NI
\emph{Case 1.} $s \not\in support(m_1)$.

Then $m_1$ is a mixed strategy $G''$, so
$r \ R \cup Q \ m_1$ holds in $G''$
by the hereditarity of $R \cup Q$.
\II

\NI
\emph{Case 2.} $s \in support(m_1)$.

If $s \: R \ m_2$ holds in $G$, then, by the regularity of $R$,
$r \: R \ m_1 [s/m_2]$ holds in $G$.  

If $s \: Q \ m_2$ holds in $G$, then by the fact that $R$ is closed under $Q$ \
$r \: R \ m_1 [s/m_2]$ holds in $G$, as well.
By assumption $m_2$ is a mixed strategy of $G''$, so $s \not\in
support(m_2)$ and consequently $s \not\in support(m_1 [s/m_2])$.  So
by the first clause of the regularity condition for some mixed
strategy $m_3$ with $r, s \not\in support(m_3)$ we have $r \: R \ 
m_3$.
Hence, by Case 1, $r \ R \cup Q \ m_1$ holds in $G''$.
\II

This proves that $G'' \Ra_{\hspace{-1mm} \: 1, R \cup Q} G' \cap G''$.
To show that $G' \Ra_{\hspace{-1mm} \: 1, R \cup Q} G' \cap G''$
we again consider two cases.
\II

\NI
\emph{Case 1.} $r \not\in support(m_2)$.

Then $m_2$ is a mixed strategy $G'$, 
so $s \ R \cup Q \ m_2$ holds in $G'$
by the hereditarity of $R \cup Q$.
\II

\NI
\emph{Case 2.} $r \in support(m_2)$.

Recall that $s \: R \cup Q \ m_2$ holds in $G$.
If $s \: R \ m_2$ holds in $G$, then, by the regularity of $R$, $s \: R \ 
m_2[r/m_1]$ holds in $G$.  If $s \: Q \ m_2$ holds in $G$, then by the fact
that $R$ is closed under $Q$ \  $s \: R \ m_2[r/m_1]$
holds in $G$, as well.
By assumption $m_1$ is a mixed strategy of $G'$, so $r \not\in support(m_1)$
and consequently $r \not\in support(m_2[r/m_1])$.
So $m_2[r/m_1]$ is a mixed strategy of $G'$.
This shows $G' \myra_{\hspace{-1mm} R \: } G' \cap
G''$.

By the Equivalence Lemma \ref{lem:equm} the relations
$\myra_{\hspace{-1mm} R}$ and $\Ra_{\hspace{-1mm} R}$ coincide, so
some mixed strategy $m_3$ of $G' \cap G''$
exists such that $s \: R \ m_3$ and
a fortiori $s \: R \cup Q \ m_3$, holds in $G'$.
\II

This proves that $G' \Ra_{\hspace{-1mm} \: 1, R \cup Q} G' \cap G''$.
\HB
\VV

This result can be directly applied to the combination of the
elimination by strict dominance by mixed strategies and by the randomized
redundance.  Indeed, we already noticed that both mixed dominance
relations are hereditary, so their union is, as well.  Also, we
already saw that strict dominance by means of mixed strategies is
regular and it is easy to see it is closed under the randomized
redundance. So by the Redundance
Elimination Theorem \ref{thm:elim2}$(i)$
and the above result we can draw the following
conclusions.

\begin{theorem}[Combined Mixed Strict Elimination] \label{thm:mcombined}
The  $\Ra_{\hspace{-1mm} SM \cup {\it PDM} \: }$ relation is $\sim$-UN.
\HB
\end{theorem}

\section{Combining Nice Weak Dominance with \mbox{\qquad} Randomized Redundance}
\label{sec:mnweak}

Finally, we provide a proof of another result of \cite{MS97} that deals with the
nice weak dominance by mixed strategies. 
This concept is obtained by generalizing in the obvious way the definition
of nice weak dominance to the case when the dominating strategy is mixed.

Recall from Section \ref{sec:borm} that given a game $G$ we write
$s''_i \ {\it WM} \ m'_i$ when the strategy $s''_i$ is weakly
dominated in $G$ by the mixed strategy $m'_i$. We also write $s''_i \ 
{\it NWM} \ m'_i$ when the strategy $s''_i$ is nicely weakly dominated
in $G$ by the mixed strategy $m'_i$, that is when $s''_i \ {\it WM} \ 
m'_i$ and $s''_i$ and $m'_i$ are compatible.

As in Section \ref{sec:nweak} we summarize first the relevant properties of the
nice weak mixed dominance relation.

\begin{lemma}[Nice Mixed Weak Dominance] \label{lem:nicem}
\mbox{} \vspace{-3mm} 
  \begin{enumerate} \smallromani

\item ${\it NWM}$ is regular.

\item ${\it NWM}$ is closed under {\it PEM}.

\item ${\it NWM} \cup {\it PEM}$ is hereditary.
  \end{enumerate}
\end{lemma}

\Proof Fix  a game $(S_1, \LL, S_n, p_1, \LL,p_n)$. 

\NI
$(i)$ 
Suppose that for some $\alpha \in (0,1]$ and some strategy $s$ 
and a mixed strategy $m$ of player $i$ \ 
\[
s \ {\it NWM} \ (1 - \alpha) s + \alpha \: m
\]
holds. By definition for all $j \in [1..n]$ and
all $s_{-i} \in S_{-i}$
\[
p_j((1 - \alpha) s + \alpha \: m, s_{-i}) = (1 - \alpha) p_j(s, s_{-i}) +  \alpha \: p_j(m, s_{-i}),
\]
so for all $op \in \{=, <, \leq\}$
\[
\mbox{$p_j(s, s_{-i}) \: op \: p_j((1 - \alpha) s + \alpha \: m, s_{-i})$ iff $p_j(s, s_{-i}) \: op \: p_j(m, s_{-i})$.}
\]
This implies $s \ {\it NWM} \ m$. 

Next, consider the strategies $t_1$ and $t_2$ and mixed
strategies $m_1$ and $m_2$ of player $i$.
For some $\alpha \in [0,1]$ and a mixed strategy $m$ we have
$m_1 = \alpha \: t_2 + (1 - \alpha) m$.
By definition for all $j \in [1..n]$ and all $s_{-i} \in S_{-i}$

\begin{equation}
  \label{eq:mone}
p_j(m_1, s_{-i}) =  \alpha \: p_j(t_2, s_{-i}) + (1 - \alpha) p_j(m, s_{-i})
\end{equation}
and
\begin{equation}
  \label{eq:mtwo}
p_j(m_1 [t_2/m_2], s_{-i}) =  \alpha \: p_j(m_2, s_{-i}) + (1 - \alpha) p_j(m, s_{-i}).  
\end{equation}
It is now easy to check that 
$t_1 \ {\it WM} \ m_1$ and $t_2 \ {\it WM} \ m_2$ implies
$t_1 \ {\it WM} \ m_{1}[t_2/m_2]$.

Suppose now that $t_1 \ {\it WNM} \ m_1$ and $t_2 \ {\it WNM} \ m_2$.
We prove that $t_1$ and $m_{1}[t_2/m_2]$ are compatible.
So suppose that for some $i \in [1..n]$ and $s_{-i} \in S_{-i}$
\[
p_i(t_1, s_{-i}) = p_i(m_1 [t_2/m_2], s_{-i}).
\]
Then by (\ref{eq:mone}) and (\ref{eq:mtwo}) and the fact
that $t_1 \ {\it WM} \ m_1$ and $t_2 \ {\it WM} \ m_2$
\[
\mbox{$p_i(t_1, s_{-i}) = p_i(m_1, s_{-i})$ and $p_i(t_2, s_{-i}) = p_i(m_2, s_{-i})$.}
\]
So by the compatibility of $t_1$ and $m_1$ and of
$t_2$ and $m_2$ for all $j \in [1..n]$
\[
\mbox{$p_j(t_1, s_{-i}) = p_i(m_1, s_{-i})$ and $p_j(t_2, s_{-i}) = p_i(m_2, s_{-i})$,}
\]
so again by (\ref{eq:mone}) and (\ref{eq:mtwo})
\[
p_j(t_1, s_{-i}) = p_j(m_1 [t_2/m_2], s_{-i}).
\]
$(ii)$ 
The proofs of the relevant two properties of ${\it NWM}$ are
analogous to the proof of $(i)$ and are omitted. 
\III

\NI
$(iii)$ Analogous to the proof of the Nice Weak Dominance Lemma
\ref{lem:nice}$(iii)$ and omitted.
\HB
\VV

We can now apply to nice weak mixed dominance the
Inherent Mixed Elimination Theorem \ref{thm:strict4}. This way we obtain the following
result.

\begin{theorem}[Inherent Nice Weak Mixed Elimination] \label{thm:nweakm-inh}
\mbox{} \vspace{-3mm} 
  \begin{enumerate} \smallromani

\item 
The $\Ra_{\hspace{-1mm} inh-NWM}$ relation is UN.

\item 
The  $\Ra_{\hspace{-1mm} inh-NWM}$ relation satisfies the one-at-a-time property.
\HB
\end{enumerate}
\end{theorem}

Further, on the account of the Redundance Elimination Theorem
\ref{thm:elim2}$(i)$ for $R := \emph{NWM}$ and $Q :=\emph{PEM}$ 
all assumptions of
the Combination Theorem \ref{thm:mcombination} are satisfied. We 
can then draw the following conclusion.

\begin{theorem}[Nice Weak Mixed Elimination] \label{thm:mnweak1}
The  $\Ra_{\hspace{-1mm} \it{NWM}  \cup \it{PEM}}$
relation is $\sim$-UN.
\HB
\end{theorem}

To draw a similar conclusion for the weak dominance by mixed strategies, 
as in Section \ref{sec:nweak} we provide three alternative conditions.
The first one, TDIM, is the direct counterpart of the TDI condition 
(\ref{eq:tdi}):

\[
  \begin{array}{l}
\mbox{for all $i, j \in [1..n]$, $r_i \in S_i$, $m_i \in M_i$  and $s_{-i} \in S_{-i}$} \\
\mbox{$p_{i}(r_i, s_{-i}) = p_{i}(m_i, s_{-i})$ implies $p_{j}(r_i, s_{-i}) = p_{j}(m_i, s_{-i})$}
  \end{array}
\]
Equivalently, for all $i \in [1..n]$, $r_i \in S_i$ and $m_i \in M_i$,
$r_i$ and $m_i$ are compatible.

Indeed, the compatibility as a mixed dominance relation is hereditary, 
so the TDIM condition implies that nice weak dominance and weak dominance,
both by mixed strategies, coincide on all restrictions.

The second one, TDIM$^+$, is the counterpart of the TDI$^+$ condition of Section
\ref{sec:nweak}. Given an initial game $H$ we postulate that

\[
  \begin{array}{l}
\mbox{for all restrictions $G := (S_1, \LL, S_n, p_1, \LL, p_n)$ of $H$,} \\
\mbox{for all $i \in [1..n]$, $r_i \in S_i$ and $m_i \in M_i$} \\
\mbox{if $m_i$ weakly dominates $r_i$ in $G$, then $r_i$ and $m_i$ are compatible in $G$.}
  \end{array}
\]

Then each time an $\Ra_{\hspace{-1mm} {\it WM}}$ reduction can take
place, it is in fact an $\Ra_{\hspace{-1mm} {\it NWM}}$ reduction.
The last alternative, TDI$^*$, was proposed in  \cite{MS97}. It refers to the notion
of the very weak dominance introduced in Section \ref{sec:nweak},
now used as a mixed dominance relation:

\[
  \begin{array}{l}
\mbox{for all restrictions $G := (S_1, \LL, S_n, p_1, \LL, p_n)$ of $H$,} \\
\mbox{for all $i \in [1..n]$, $r_i \in S_i$ and $m_i \in M_i$} \\
\mbox{if $m_i$ very weakly dominates $r_i$ in $G$, then} \\
\mbox{either $m_i$ weakly dominates $r_i$ in $G$ or $r_i$ and $m_i$ are payoff equivalent in $G$.}
  \end{array}
\]

Then the following result holds.

\begin{theorem}[Weak Mixed Elimination] \label{thm:mweak1}
Consider a game $H$ that satisfies the TDI$\: ^*$ condition.
Then the  $\Ra_{\hspace{-1mm} \it{WM}  \cup \it{PEM}}$ relation is $\sim$-UN.
\end{theorem} 
\Proof
We proceed as in Section \ref{sec:nweak}. 
Denote the very weak mixed dominance relation by $VWM$. Note that
\begin{itemize}
\item $WM \sse VWM$,

\item $VWM$ is hereditary.

\end{itemize}

Additionally, by the TDI$^{*}$ assumption,

\begin{itemize}
\item $VWM \sse WM \cup {\it PEM}$

\end{itemize}
holds in all restrictions of the initial game $H$.

So under the TDI$^{*}$ assumption $WM \cup {\it PEM}$ is
hereditary since ${\it PEM}$ is hereditary.
By the Combination Theorem \ref{thm:mcombination} we
conclude that the $\Ra_{\hspace{-1mm} \it{WM} \cup \it{PEM}}$
relation is $\sim$-UN.

To establish another form of order independence 
involving nice mixed weak dominance
we need the following observation.

\begin{note}[Left Commutativity] \label{not:left2} 
\mbox{} \vspace{-3mm} 
  \begin{enumerate} \smallromani

\item $\Ra_{\hspace{-1mm} {\it PEM} \: }$ left commutes with $\Ra_{\hspace{-1mm} {\it NWM}}$.

\item $\Ra_{\hspace{-1mm} {\it PEM} \: }$ left commutes with $\Ra_{\hspace{-1mm} {\it WM}}$.
  \end{enumerate}
\end{note}
\Proof 
$(i)$
By the Nice Mixed Weak Dominance Lemma \ref{lem:nicem}$(ii)$
${\it NWM}$ is closed under the randomized redundance. 
The rest of the proof is now analogous to the proof of the 
Left Commutativity Note \ref{not:left1} and is omitted. 
\III

\NI
$(ii)$ By the same argument as in $(i)$.
\HB
\VV

As in Section \ref{sec:nweak} we can now draw the following results due to \cite{MS97}.

\begin{theorem}[Structured Nice Weak Mixed Elimination] \label{thm:weak4}
  Suppose 

\NI
that $G \Ra^{\hspace{-1mm} *}_{\hspace{-1mm} {\it NWM} \: } G'$
  and $G \Ra^{\hspace{-1mm} *}_{\hspace{-1mm} {\it NWM} \: } G''$, where
  both $G'$ and $G''$
are closed under the $\Ra_{\hspace{-1mm} {\it NWM}}$ reduction.

Then for some $\sim$-equivalent games $H'$ and $H''$
closed under the $\Ra_{\hspace{-1mm} {\it NWM} \cup {\it PEM}}$ reduction
we have $G' \Ra^{\hspace{-1mm} *}_{\hspace{-1mm} {\it PEM} \: } H'$ and 
$G'' \Ra^{\hspace{-1mm} *}_{\hspace{-1mm} {\it PEM} \: } H''$.
\HB
\end{theorem}

\begin{corollary}[Structured Weak Mixed Elimination] \label{cor:weak4}
Consider a game $G$ that satisfies the TDI$\: ^{*}$ condition.
Suppose that $G \Ra^{\hspace{-1mm} *}_{\hspace{-1mm} WM \: } G'$ and 
$G \Ra^{\hspace{-1mm} *}_{\hspace{-1mm} WM \: } G''$, where both $G'$ 
and $G''$ are closed under the $\Ra_{\hspace{-1mm} WM}$ reduction.

Then for some $\sim$-equivalent games $H'$ and $H''$
closed under the $\Ra_{\hspace{-1mm} {\it WM} \cup {\it PEM}}$ reduction
we have $G' \Ra^{\hspace{-1mm} *}_{\hspace{-1mm} {\it PEM} \: } H'$ and 
$G'' \Ra^{\hspace{-1mm} *}_{\hspace{-1mm} {\it PEM} \: } H''$.
\HB
\end{corollary}

\section{Conclusions}
\label{sec:conclusions}

In this paper we presented uniform proofs of order independence for
various strategy elimination procedures. The main ingredients of our approach
were reliance on Newman's Lemma and related results on the abstract
reduction systems, and an
analysis of the structural properties of the dominance relations.
This exposition allowed us to clarify which structural properties
account for the order independence of the entailed reduction relations
on the games.

In Figure \ref{fig:summary} below we summarize the order independence
results discussed in this article.  We use here the already introduced
abbreviations, so:
\begin{description}
\item[--]
\emph{S} denotes strict dominance, 
\item[--]
\emph{W} denotes weak dominance, 
\item[--]
\emph{NW} denotes nice weak dominance, 
\item[--]
\emph{PE} denotes payoff equivalence.
\end{description}
Further, \emph{RM} stands for the `mixed strategy' version of the dominance relation $R$
and \emph{inh-R} stands for the `inherent' version of the (mixed) dominance relation $R$
discussed in Sections \ref{sec:bor} and \ref{sec:borm}.

Recall also that UN stands for the uniqueness of the normal form, i.e., for the order independence
and $\sim$-UN is its `up to the game equivalence' version.
All the results refer to the order independence of the
$\Ra_{\hspace{-1mm} R}$ reduction relation on games,
introduced in Section \ref{sec:dominance}.

\begin{figure}[htbp]
  \begin{center}
\begin{tabular}{|l|r|r|l|}
\hline
Dominance                         & Property & Proved                    & Result originally due to \\
Notion                             &                  & in Section                & \\
\hline \hline
\emph{S}                           & UN               & \ref{sec:dominance}    & \cite{GKZ90}, \\
                                   &                  &                        & \cite{Ste90} \\
$inh-W$                       & UN               & \ref{sec:bor}          & \cite{Bor90} \\
$inh-NW$                      & UN               & \ref{sec:nweak}    & \\
\emph{SM}                          & UN               & \ref{sec:mixeddom}    & \cite{OR94} \\
$inh-WM$                      &        UN        & \ref{sec:borm}    & (\cite{Bor90}: equal to \emph{SM}) \\
$inh-NWM$                   & UN               & \ref{sec:mnweak}    & \\
\emph{PE}                     & $\sim$-UN        & \ref{sec:pure-equ}    & \\
$\emph{S} \cup \emph{PE}$         & $\sim$-UN        & \ref{sec:combining}    &  \\
$\emph{NW} \cup \emph{PE}$         & $\sim$-UN        & \ref{sec:nweak}    & \cite{MS97} \\
\emph{PEM}                    & $\sim$-UN        & \ref{sec:randomized}    & \\
$\emph{SM} \cup \emph{PEM}$        & $\sim$-UN       & \ref{sec:combiningm}    &  \\
$\emph{NWM} \cup \emph{PEM}$        & $\sim$-UN       & \ref{sec:mnweak}    & \cite{MS97} \\
\hline
\end{tabular}
    \caption{Summary of the order independence results}
    \label{fig:summary}
  \end{center}
\end{figure}

The reduction relations on games that we studied are naturally
related.  For example we have $\Ra_{\hspace{-1mm} S} \sse
\Ra_{\hspace{-1mm} SM}$, with the strict inclusion for some games.
However, the respective results about these reduction relations are
not related.  For example, the fact that $\Ra_{\hspace{-1mm} S}$ is
UN not a special case of the fact that $\Ra_{\hspace{-1mm} SM}$ is
UN.

Indeed, given two abstract reduction systems $(A, \myra_{\hspace{-1mm}
  1 \: })$ and $(A, \myra_{\hspace{-1mm} 2 \: })$ such that
$\myra_{\hspace{-1mm} 1 \: } \sse \myra_{\hspace{-1mm} 2 \: }$ the
uniqueness of a normal form with respect to $\myra_{\hspace{-1mm} 2 \:
  }$ does not imply the uniqueness of a normal form with respect to
$\myra_{\hspace{-1mm} 1}$. Indeed, just take $\myra_{\hspace{-1mm} 1
  \: } := \{(a,b), (a,c)\}$ and $\myra_{\hspace{-1mm} 2 \: } :=
\myra_{\hspace{-1mm} 1 \: } \cup \{(b,d), (c,d)\}$. This example also
shows that weak confluence of $\myra_{\hspace{-1mm} 2}$ does not imply
weak confluence of $\myra_{\hspace{-1mm} 1}$.  So the weak confluence
of $\Ra_{\hspace{-1mm} S}$ is not a consequence of the weak
confluence of $\Ra_{\hspace{-1mm} SM}$.  The same remarks apply to
other pairs of dominance relations.

The provided proofs of the order independence results break down for
infinite games.  The reason is that the crucial assumption of Newman's
Lemma, namely that that no infinite $\myra$ sequences exist, does not
hold then anymore.  Moreover, for infinite games the Equivalence Lemma
\ref{lem:equ1} does not hold.  Still, it would be interesting to try
to establish the main result of \cite{DS02} using the abstract
reduction systems techniques.


\subsection*{Acknowledgement}
We thank Bernhard von Stengel for a useful email exchange on the
subject of mixed strategies in the initial stage of this research,
Femke van Raamsdonk for her comments on abstract reduction systems,
and Jeroen Swinkels and the referees for helpful suggestions.


\bibliographystyle{handbk}



\end{document}